\documentclass[fleqn,useAMS,usenatbib]{mn2e}
\usepackage{color}

\usepackage{exscale}
\usepackage{graphicx}
\usepackage{natbib}
\usepackage{rotating}
\usepackage{rotate}
\usepackage{afterpage}
\usepackage{fancyhdr}	
\usepackage{amsmath}
\usepackage{amssymb}
\usepackage{booktabs,threeparttable}
\usepackage{relsize}
\usepackage{lscape}
\usepackage{epsfig}
\usepackage{tabularx}
\usepackage{longtable}
\usepackage{ltxtable}
\usepackage{txfonts}
\usepackage{natbib}
\usepackage{bbm}
\usepackage[outercaption]{sidecap}
\usepackage{afterpage}
\usepackage{widetext}

\def\apj{ApJ}%
\def\apjl{ApJ}%
\def\aap{A\&A}%
\def\aaps{A\&AS}%
\def\mnras{MNRAS}%
\def\prd{Phys.~Rev.~D}%
\def\jcap{JCAP}%
\def\physrep{Phys.~Rep.}%
\newcommand{\tbf}{\textbf}
\newcommand{\ti}{\textit}

\newcommand{\bea}{\begin{eqnarray}}
\newcommand{\be}{\begin{equation}}
\newcommand{\ben}{\begin{enumerate}}
\newcommand{\bi}{\begin{itemize}}
\newcommand{\eea}{\end{eqnarray}}
\newcommand{\ee}{\end{equation}}
\newcommand{\ei}{\end{itemize}}
\newcommand{\een}{\end{enumerate}}

\newcommand{\matC}{\mathbf C}

\newcommand{\like}{L}
\newcommand{\prob}{P}
\newcommand{\probr}{P_r}

\newcommand{\pco}{\vek p_\mr{c}}
\newcommand{\pnu}{\vek p_\mr{n}}

\newcommand{\D}{\vek D}

\newcommand{\M}{\vek M}

\newcommand{\om}{\Omega_\mr m}
\newcommand{\omb}{\Omega_\mr b}
\newcommand{\sig}{\sigma_8}
\newcommand{\ns}{n_s}
\newcommand{\w}{w_0}
\newcommand{\wa}{w_a}

\renewcommand{\d}{{\rm d}}
\newcommand{\pd}{P_{\delta}}

\newcommand{\eps}{\epsilon}

\newcommand{\mr}{\mathrm}

\renewcommand{\d}{{\rm d}}

\newcommand{\fred}{f_\mr{red}}

\def\vek{\mathbf}

\title[The impact of intrinsic alignment on current and future cosmic shear surveys]{The impact of intrinsic alignment on current and future cosmic shear surveys}
\author[Krause et al.]
{\parbox{\textwidth}{Elisabeth Krause$^{1,2}$,\thanks{E-mail: \texttt{aelisabeth.krause@gmail.com}}
Tim Eifler$^{3,2}$,
Jonathan Blazek$^{4}$\vspace{0.4cm}}\\
\parbox{\textwidth}{$^{1}$ Kavli Institute for Particle Cosmology and Astrophysics, Stanford University, Stanford, CA 94305, USA\\
$^{2}$ Department of Physics and Astronomy, University of Pennsylvania, Philadelphia, PA 19104, USA \\
$^{3}$ Jet Propulsion Laboratory, California Institute of Technology, 4800 Oak Grove Dr., Pasadena, CA 91109\\
$^{4}$ Center for Cosmology and AstroParticle Physics, The Ohio State University, 191 W. Woodruff Ave., Columbus, OH 43210, USA\\
}}

\begin{document}

\date{accepted received}

\maketitle

\label{firstpage}

\begin{abstract}
Intrinsic alignment (IA) of source galaxies is one of the major astrophysical systematics for ongoing and future weak lensing surveys. Several IA models have been proposed in the literature and constrained through observations. Their impact on cosmological constraints has been examined using the Fisher Information Matrix in conjunction with approximate covariance computation schemes.\\
This paper presents the first forecasts of the impact of IA on cosmic shear measurements for future surveys (DES, Euclid, LSST, WFIRST) using simulated likelihood analyses and realistic covariances that include higher-order moments of the density field in the computation. We consider a range of possible IA scenarios and test mitigation schemes, which parameterize IA by the fraction of red galaxies, normalization, luminosity and redshift dependence of the IA signal (for a subset we consider joint IA and photo-z uncertainties). \\ 
Compared to previous studies we find smaller biases in time-dependent dark energy models if IA is ignored in the analysis; the amplitude and significance of these biases vary as a function of survey properties (depth,  statistical uncertainties), luminosity function, and IA scenario: Due to its small statistical errors and relatively shallow observing strategy Euclid is significantly impacted by IA. LSST and WFIRST benefit from their increased survey depth, while the larger statistical errors for DES decrease IA's relative impact on cosmological parameters.  \\
The proposed IA mitigation scheme removes parameter biases due to IA for DES, LSST, and WFIRST even if the shape of the IA power spectrum is only poorly known; successful IA mitigation for Euclid requires more prior information. We explore several alternative IA mitigation strategies for Euclid; in the absence of alignment of blue galaxies we recommend the exclusion of red (IA contaminated) galaxies in cosmic shear analyses. We find that even a reduction of 20\% in the number density of galaxies only leads to a 4-10\% loss in cosmological constraining power.
\end{abstract}


\section{Introduction}
\label{sec:intro}
Weak gravitational lensing by large-scale structure, so-called cosmic shear, is a promising technique to constrain cosmological parameters. Cosmic shear is among the key science cases of various ongoing surveys, such as Kilo-Degree Survey (KiDS\footnote{http://www.astro-wise.org/projects/KIDS/}), Hyper Suprime Cam (HSC\footnote{http://www.naoj.org/Projects/HSC/HSCProject.html}), and Dark Energy Survey (DES\footnote{www.darkenergysurvey.org/}). Independent of any assumptions about the relationship between dark and luminous matter, it provides valuable information on both the geometry and structure growth of the Universe \citep{hyg02,wmh05,jjb06,shj10,lds12,hwm12,heh14}.

If systematics can be sufficiently controlled for future missions, such as the Large Synoptic Survey Telescope (LSST\footnote{http://www.lsst.org/lsst}), Euclid\footnote{sci.esa.int/euclid/} and the Wide-Field Infrared Survey Telescope (WFIRST\footnote{http://WFIRST.gsfc.nasa.gov/}), cosmic shear has the potential to be the most constraining cosmological probe. If systematics control does not improve, lensing constraints will be substantially weaker than constraints from other probes, such as SN1a, Baryonic Acoustic Oscillations, and redshift-space distortions \citep{wme13}.

Important systematic effects that complicate the extraction of cosmological information from cosmic shear are shear calibration \citep{hir03,hkm05} and photo-z errors \citep{mhh06,beh10,hzm10}, baryonic effects \citep{jzl06,zrh08,shs11,zsd13,shs13,ekd14} and intrinsic alignment (IA) of source galaxies.

Cosmic shear is typically measured through two-point correlations of observed galaxy ellipticities. In the weak lensing regime, the observed ellipticity of a galaxy is the sum of intrinsic ellipticity $\epsilon^{\rm I}$ and gravitational shear $\gamma$: $\epsilon^{\rm{obs}} \approx \epsilon^{\rm I} +\gamma$. If the intrinsic shapes of galaxies are not random, but spatially correlated, these intrinsic alignment correlations can contaminate the gravitational shear signal and lead to biased measurements if not properly removed or modeled. Since early work establishing its potential effects \citep{crm00,hrh00,ckb01,cnp01}, IA has been examined through observations \citep[e.g.,][]{hmi07,jma11,bms12,smm14}, analytic modeling, and simulations  \citep[e.g.,][]{sfc12, tsm14,tmd14} - see \cite{tri14}, \cite{Joachimi15}, and references therein, for recent reviews. A fully predictive model of IA would include the complex processes involved in the formation and evolution of galaxies and their dark matter halos, as well as how these processes couple to large-scale environment. In the absence of such knowledge, analytic modeling of IA on large scales relates observed galaxy shapes to the gravitational tidal field, and typically considers either tidal (linear) alignments, or tidal torquing models.

The shapes of elliptical, pressure supported galaxies are often assumed to align with the surrounding dark matter halos, which are themselves aligned with the stretching axis of the large-scale tidal field \citep{ckb01,his04}. This tidal alignment model leads to shape alignments that scale linearly with fluctuations in the tidal field, and it is thus sometimes referred to as ``linear alignment,'' although nonlinear contributions may still be included \citep{brk07,bms11,bvs15}. For spiral galaxies, where angular momentum is likely the primary factor in determining galaxy orientation, IA modeling is typically based on tidal torquing theory, leading to a quadratic dependence on tidal field fluctuations \citep{pls00,ckb01,huz02,lep08}, although on sufficiently large scales, a contribution linear in the tidal field may dominate. Due to this qualitative difference in assumed alignment mechanism, source galaxies are often split by color into ``red'' and ``blue'' samples, as a proxy for elliptical and spiral types. Indeed, blue samples consistently exhibit weaker IA on large scales, supporting the theory that tidal alignment effects are less prominent in spirals \citep{flw09,hmi07,mbb11}. On smaller scales, IA modeling must include a one-halo component to describe how central and satellite galaxies align with each other and with respect to the distribution of dark matter \citep{scb10}. Numerical simulations, especially those including hydrodynamical physics, have recently become powerful tools for constructing these models \citep{sfc12, jsh13,jsb13, tsm14,tmd14}. However, due to the cumulative uncertainty in modeling the cosmic shear signal on angular scales corresponding to the one-halo term regime, e.g. from uncertainties in modeling the non-linear matter power spectrum \citep[e.g.,][]{heh09,e11} and the impact of baryons \citep[e.g.,][]{dsb11,zsd13}, a more conservative analysis strategy might be to exclude corresponding scales from the analysis \citep[e.g.][]{laa11}.

In this work, we focus on IA for red source galaxies on scales outside the one-halo regime, testing the potential impact on cosmic shear analyses using several different tidal alignment scenarios. Given recent observational results, IA for blue source galaxies is likely subdominant. Although we examine the potential effects of tidal alignment for blue galaxies, we leave a detailed treatment of IA contamination from blue galaxies for future work. 

\section{Intrinsic Alignment Modeling}
\label{sec:iabasics}

We consider a class of shape alignment models for which the correlated component of galaxy shapes is proportional to the gravitational tidal field. These models are often referred to as ``linear alignment'' or ``tidal alignment,'' and are most frequently used to describe IA of elliptical (red) galaxies. In this work, we do not use a separate model for the (much weaker) alignment of blue galaxies. For further discussion, see \cite{ckb01,his04,jma11,bms11,bvs15}.

The underlying principle of tidal alignment models is that the gravitational collapse of an initially spherical overdensity in a tidal gravitational field leads to triaxial haloes, such that the halo will be prolate if the overdensity is stretched by the large scale tidal field and oblate if it is compressed \citep{ckb01}. Under the assumption that the shape and orientation of an elliptical galaxy are determined by the shape of the halo in which it resides, this mechanism leads to a net correlation of intrinsic galaxy ellipticities with the gravitational tidal field. The lowest-order contribution is a linear relationship to the tidal field, which can be expressed in Fourier space in terms of the density contrast $\delta$:
\be
\epsilon^{\rm{IA}}_{+,\times}(\mathbf k, z) = -A(z)\frac{(k_x^2 - k_y^2, 2 k_x k_y)}{k^2}\mathcal{S}\left[\delta(\mathbf k,z_{\rm IA})\right] \,,
\ee
where the ellipticity components, denoted $(+,\times)$ are measured with respect to the $x$-axis. The amplitude $A(z)$ depends on redshift and galaxy properties, and $\mathcal{S}$ is a filter that smooths the tidal field on the relevant scale \citep[see][]{bvs15}. The redshift at which IA is set, $z_{IA}$, the amplitude $A(z)$, and the treatment of the density field can differ in various implementations of the tidal alignment model, as detailed below. The intrinsic alignments contribute to the observed ellipticity correlations through through both their auto-correlation and their cross-correlation with the density field, known respectively as the ``II'' and ``GI'' terms \citep[see][]{his04}. At leading order, gravitational lensing produces $E$-mode (curl-free) shape correlations, and we are thus concerned with the $E$-mode IA contribution.\footnote{$B$-modes can be produced by IA contamination as well as higher-order lensing effects.} For simplicity, we introduce the following notation for the $E$-mode component of the intrinsic ellipticity field:
\be
\epsilon^{\rm{IA}}_{E} = -A(z) \mathcal{E}(k,z) \, .
\ee
The relevant $3D$ IA power spectra can then be written:
\bea
P_{\mathrm{II}}(k,z) = A^2(z) P_{\mathcal{E}\mathcal{E}}(k,z)\, , \\
P_{\mathrm{G I}}(k,z) = -A(z) P_{\delta \mathcal{E}}(k,z)\,.
\eea

\subsection{IA amplitude}
\label{sec:Aia}
The redshift at which the alignment is set, $z_{\rm{IA}}$, and the subsequent IA evolution are determined by the astrophysical processes involved in galaxy formation and evolution. It is sometimes assumed that $z_{\rm{IA}}$ is during matter domination when a halo first forms (``primordial alignment''). However, late-time accretion and mergers could have a significant impact on IA, in which case the relevant $z_{\rm{IA}}$ could be closer to the observed redshift, $z_{\rm{obs}}$. In the case of instantaneous alignment ($z_{\rm{IA}}=z_{\rm{obs}}$),
\begin{align}
A^{\rm{inst}}(z) = C_1^{\rm{inst}} \rho_{\rm m,0}(1+z)\,,
\end{align}
with a proportionality constant $C_1$ describing the strength of galaxy shape alignments, which may depend of galaxy properties and needs to be determined empirically \citep[see][for further discussion on $A(z)$]{bvs15}.

In the ``primordial alignment'' assumption, IA is set at some early time and does not evolve. In this case, since the relevant density field is determined at a redshift where nonlinear growth can be neglected, we can instead evaluate it at the observed redshift and absorb the linear growth factor, $D(z)$ into the overall amplitude:
\begin{align}
\label{eq:general_A}
 A^{\rm{early}}(z)&= C_1^{\rm{early}}\rho_{\rm cr}\Omega_{\rm m,0}(1+z_{\rm{IA}})\frac{D(z_{\rm{IA}})}{D(z)} =\frac{C_1^{\rm{early}}\rho_{\rm m,0}M}{D(z)}\,,
\end{align}
 where in the second step we have assumed that alignment happens during matter domination such that  $M \equiv (1+z_{\rm{IA}})D(z_{\rm{IA}})$ is constant. The resulting expression has no dependence on $z_{\rm{IA}}$. 
 
In practice, the redshift evolution of intrinsic alignments may be between these scenarios, and the effective amplitude will depend on the full luminosity distribution of the (red) galaxy sample of interest. In this analysis, we use the redshift and luminosity scaling
\be
\label{eq:A_L}
A(L,z)  = \frac{C_1\rho_{\rm m,0}}{D(z)}A_0 \left(\frac{L}{L_0}\right)^\beta \left(\frac{1+z}{1+z_0}\right)^{\eta}
\ee
as baseline model. In this expression $C_1 \, \rho_\mr{cr}\approx0.0134$ is a normalization derived from SuperCOSMOS observations \citep{his04, brk07}, and for the fit parameters $A,\eta_\mr{other},\beta$ we adopt the constraints from the MegaZ-LRG + SDSS LRG sample in \cite{jma11}, i.e. $A_0=5.92^{+0.77}_{-0.75}$, $\eta=-0.47^{+0.93}_{-0.96}$, and $\beta=1.10^{+0.29}_{-0.30}$, with $z_0 = 0.3$ the observationally-motivated pivot redshift and $L_0$ the pivot luminosity corresponding to an absolute $r$-band magnitude of $-22$.
For our implementation, we average $A(L,z)$ over the red galaxy luminosity distribution to compute the average IA amplitude for the red source galaxy population at a given redshift (see Sect.~\ref{sec:fred}), and include an additional parameter $\eta_{\rm{high-z}}$ to account for the uncertainty in the extrapolation of this redshift scaling,
\be
A(m_{\mathrm{lim}},z) = \Big\langle A(L,z)\Big\rangle_{\phi_{\rm{red}}}\times \left[\Theta(z_1 -z)+\Theta(z -z_1) \left(\frac{1+z}{1+z_1}\right)^{\eta_\mr{high-z}}\right]\,,
\ee
with $m_{\mathrm{lim}}$ the survey's limiting magnitude, and $\Theta$ the step function. The term in square brackets is a truncated power-law in $(1+z)$, which is unity for $z<= z_1$ and with slope $ \eta_{\rm{high-z}}$ for $z> z_1$. As the fiducial redshift scaling is based on the MegaZ-LRG + SDSS LRG sample with $z  \leq  0.7$, we choose $z_1 = 0.7$ and marginalize over $\eta_{\rm{high-z}}$ (see Sect.~\ref{sec:like_theory} and Tab.~\ref{tab:nuisance} for details).
\subsection{IA density power spectra models} 
\label{sec:IAmod}
Four effects can produce nonlinearities in the intrinsic shape correlations: (1) nonlinear dependence of intrinsic galaxy shape on the tidal field (e.g., quadratic tidal torquing); (2) nonlinear evolution of the dark matter density field, leading to nonlinear evolution in the tidal field; (3) a nonlinear bias relationship between the galaxy and dark matter density fields; (4) the IA field actually observed is weighted by the local density of galaxies used to trace the shapes. In this work, we assume a pure tidal alignment model, i.e.\ that intrinsic galaxy shapes depend linearly on the tidal field, even on small scales. However, nonlinearities from the other three effects must still be considered. The following versions of the tidal alignment model treat these effects differently. We also implement tidal field smoothing only in the full tidal alignment model, as described below, allowing us to test the impact of different smoothing procedures, since the physically correct smoothing is unknown.
\paragraph*{Linear Alignment (LA)}
The most basic version of tidal alignment includes only the leading order contribution to IA correlations (i.e.\ from the linear density field $\delta_{\rm lin}$):
\be
\label{eq:LA}
P_{\delta\mathcal{E}}(k,z) =  P_{\mathcal{E}\mathcal{E}}(k,z) = P_{\mathrm{lin}}(k,z)
\ee

\paragraph*{Nonlinear Alignment (NLA)}
To account for the nonlinear evolution of the density field, the linear power spectrum in Eq.~\ref{eq:LA} can be replaced with the nonlinear version:
\be
P_{\delta\mathcal{E}}(k,z) =  P_{\mathcal{E}\mathcal{E}}(k,z) = P_{\rm{NL}}(k,z)
\ee
This approach, often called the nonlinear alignment (NLA) model, was first used by \cite{his04} and more fully developed by \cite{brk07}. Due to its relative simplicity and ability to more accurately model observed intrinsic shape correlations, the NLA model has been used in weak lensing forecasts and analyses \citep[e.g.][]{hwm12}. However, the NLA model neglects nonlinearities from source density weighing and nonlinear galaxy biasing, the former of which can be similar in size or larger than the correction due to nonlinear structure growth.

\paragraph*{Freeze-in Model (FR)}
In the primordial alignment scenario, intrinsic shape correlations are set at early times, when nonlinear evolution of the density field can be largely ignored, and are thus proportional to the linear tidal field. For the GI term, however, the nonlinear growth of the density field should still be included. To capture these effects, \cite{kirk12} suggested a model in which the density-ellipticity cross-correlation was given by the geometric mean of the linear and nonlinear power spectra:
\bea
P_{\delta\mathcal{E}}(k,z) &=&  \sqrt{P_{\rm lin}(k,z) P_{\rm NL}(k,z)} \, ,\\
P_{\mathcal{E}\mathcal{E}}(k,z)&=& P_{\mathrm{lin}}(k,z) \, .
\eea
Although approximate, this approach allows the cross-correlation of the density field at different redshifts, without needing to account for the effects of advection that appear in an Eulerian perturbation theory approach (see \citep{bvs15} for further discussion). Note that we normalize all IA models to match the observed large-scale $P_{\rm{GI}}$ amplitude at the pivot redshift $z_0 =0.3$ of \citet{jma11}, hence our $A(z)$ differs from the normalization chosen in \citet{kirk12}. After accounting for the correct redshift dependence of $A(z)$, our implementation of the freeze-in $P_{\rm{II}}$ model exhibits same redshift scaling as Eq.~(8) in \cite{kirk12}, and our $P_{GI}$ expression differs by one power of the linear growth function, which is missing in their Eq.~(9).
\paragraph*{Full tidal alignment model (TA)}
Although the NLA approach improves the model fit to data, it is not fully consistent and omits important astrophysical effects. The model of \cite{bvs15} includes all terms that contribute at next-to-leading order while simultaneously smoothing the tidal field (e.g.\ at the Lagrangian radius of the host halo). The effects of weighting by the local galaxy density can be larger than the correction from the nonlinear evolution of dark matter density, especially in the case of highly biased tracers, since the leading effect scales with the linear galaxy bias. This more complete tidal alignment model is given by \citep{bvs15}:
\bea
P_{\delta\mathcal{E}}(k,z) &=& \!\!P_{\rm NL}(k,z) + \frac{58}{105}b_1\sigma_S^2 P_{\rm lin} + b_1 P_{0|0\mathcal{E}}  \, ,\\
P_{\mathcal{E}\mathcal{E}}(k,z)&=& \!\!P_{\rm NL}(k,z) + \frac{116}{105}b_1\sigma_S^2 P_{\rm lin} + 2b_1 P_{0|0\mathcal{E}} + b_1^2 P_{0\mathcal{E}|0\mathcal{E}}\, ,
\eea
where $b_1$ is the linear bias of the source sample, and $P_{0|0\mathcal{E}}$ and $P_{0\mathcal{E}|0\mathcal{E}}$ are $\mathcal{O}(P_{\rm lin}^2)$ terms that arise from weighting the intrinsic shape field by the source density. In these expressions, a Gaussian smoothing filter with scale length $k_{\rm sm} = 1.0 h^{-1}{\rm Mpc}$ is applied to the tidal field and consistently treated in the subsequent calculations. The density variance $\sigma_{S}^2$ is determined by integrating the linear power spectrum smoothed with this filter. For this IA scenario, the redshift and luminosity scalings of the amplitude are determined by the evolution of $b_1$ and $\sigma_{S}^2$, rather than applying Eq.~(\ref{eq:A_L}). The fiducial value of $A_0$ at pivot redshift $z_0$ is multiplied by $(1+58/105 b_1(z_0) \sigma_S^2(z_0))^{-1}$ to ensure that $P_{GI}$ has the correct amplitude on large scales at that redshift (where measurements provide the tightest constraints). Note that $P_\mr{II}$ will not have the same large-scale amplitude as in the other models, since the density-weighting correction is different. 
\section{Modeling weak lensing observables and covariances}
\label{sec:lensingbasics}
All simulated likelihood analyses in this paper are computed using the weak lensing modules of \textsc{CosmoLike} \citep[see][for an early version; official release paper Krause et al. 2015 in prep]{eks14}. 

\subsection{Shear tomography power spectra}
\label{sec:powerbasics}

We compute the linear power spectrum using the \cite{eh99} transfer function and model the non-linear evolution of the density field using the updated Halofit as described in \cite{tsn12}. 
Time-dependent dark energy models ($w=w_0+(1-a)\,w_a$) are incorporated following the recipe of {\sc icosmo} \citep{rak11}, which in the non-linear regime interpolates Halofit between flat and open cosmological models \citep[also see][for more details]{shj10}.

Having obtained the density power spectra we calculate the shear power spectra as
\begin{equation}
\label{eq:pdeltatopkappa}
C ^{ij} (l) = \frac{9H_0^4 \om^2}{4c^4} \int_0^{\chi_\mr h} 
\mr d \chi \, \frac{g^{i}(\chi) g^{j}(\chi)}{a^2(\chi)} \pd \left(\frac{l}{f_K(\chi)},\chi \right) \,,
\end{equation}
with $l$ being the 2D wave vector perpendicular to the line of sight, $\chi$ denoting the comoving distance, $\chi_\mr h$ is the comoving distance to the horizon, $a(\chi)$ is the scale factor, and $f_K(\chi)$ the comoving angular diameter distance (throughout set to $\chi$ since we assume a flat Universe). The lens efficiency $g^{i}$ is defined as an integral over the redshift distribution of source galaxies $n(\chi(z))$ (see Sect. \ref{sec:surveys} for details) in the $i^\mr{th}$ tomographic interval
\begin{equation}
\label{eq:redshift_distri}
g^{i}(\chi) = \int_\chi^{\chi_{\mr h}} \mr d \chi' n^{i} (\chi') \frac{f_K (\chi'-\chi)}{f_K (\chi')} \,.
\end{equation}
We split the source galaxies into five tomographic bins, chosen such that each bin contains the same number of galaxies.
\subsection{Projected intrinsic alignment power spectra}
For a given a intrinsic alignment model, we compute the corresponding projected power spectra as
\begin{align}
\label{eq:IA1}
C_\mr{II}^{ij} (l, m_{\mathrm{lim}}) &=&  \int_0^{\chi_h} \d \chi \, \frac{n^i(\chi)n^j(\chi)}{\chi^2} \, f_\mr{red}^2(m_{\mathrm{lim}},z)\, P_\mr{I I} \left( k,z,m_{\mathrm{lim}} \right) \,, \\
\label{eq:IA2}
C_\mr{GI}^{ij} (l,m_{\mathrm{lim}}) &=&  \int_0^{\chi_h} \d \chi \, \frac{n^i(\chi) \,  g^j(\chi)}{\chi^2}  \, f_\mr{red}(m_{\mathrm{lim}},z)\, P_{\delta \mr I} \left( k,z,m_{\mathrm{lim}} \right) \,,
\end{align}
where $f_{\mr{red}}$, the fraction of red galaxies at redshift $z(\chi)$ in a particular survey, is computed from the luminosity functions for all and red galaxies, as described in Sect.~\ref{sec:fred}.

\subsection{Shear covariances}
\label{sec:covbasics}
The non-Gaussian covariance of tomographic shear power spectra $C_\kappa^{ij}(l)$ can be written as \citep{CH01,huj04,sht09}
\begin{widetext}
\bea
\label{eq:covC}
\nonumber \mr{Cov}\left( C_\kappa^{ij} (l_1), C_\kappa^{kl} (l_2) \right) &=&\frac{2\pi\delta_{l_1 l_2}}{\Omega_{\mr s} l_1 \Delta l_1}  \left[\left(C_\kappa^{ik}(l_1)+ \delta_{ik} \frac{\sigma_\eps^2}{2n^{i}_{\rm{gal}}}\right) \left(C_\kappa^{jl}(l_1)+ \delta_{jl} \frac{\sigma_\eps^2}{2n^{j}_{\rm{gal}}}\right) +\left(C_\kappa^{il}(l_1)+ \delta_{il} \frac{\sigma_\eps^2}{2n^{i}_{\rm{gal}}}\right) \left(C_\kappa^{jk}(l_1)+ \delta_{jk} \frac{\sigma_\eps^2}{2n^{j}_{\rm{gal}}}\right) \right]\\
&& +\int_{|\mathbf l|\in l_1}\frac{d^2\mathbf l}{A(l_1)}\int_{|\mathbf l'|\in l_2}\frac{d^2\mathbf l'}{A(l_2)} \left[\frac{1}{\Omega_{\mr s}}T_{\kappa,0}^{ijkl}(\mathbf l,-\mathbf l,\mathbf l',-\mathbf l') + T_{\kappa,\rm{HSV}}^{ijkl}(\mathbf l,-\mathbf l,\mathbf l',-\mathbf l') \right]\,,
\eea
\end{widetext}
\noindent with $\Omega_{\rm s}$ the survey area, $n^i_{\rm{gal}}$ the number of source galaxies in tomography bin $i$, $\sigma_\epsilon$ the ellipticity dispersion, $A(l_i) = \int_{|\mathbf l|\in l_i}d^2\mathbf l \approx 2 \pi l_i\Delta l_i$ the integration area associated with a power spectrum bin centered at $l_i$ and width $\Delta l_i$, and $T_{\kappa,0}$ and $T_{\kappa,\rm{HSV}}$ the convergence trispectrum in the absence of finite volume effects and the super-sample variance contribution to the trispectrum \citep{sht09,takada2013}. We use a halo model implementation of these terms as described in \citep{ekd14}.
\section{Survey Parameters}
\label{sec:surveys}
\begin{figure}
\includegraphics[width=8cm]{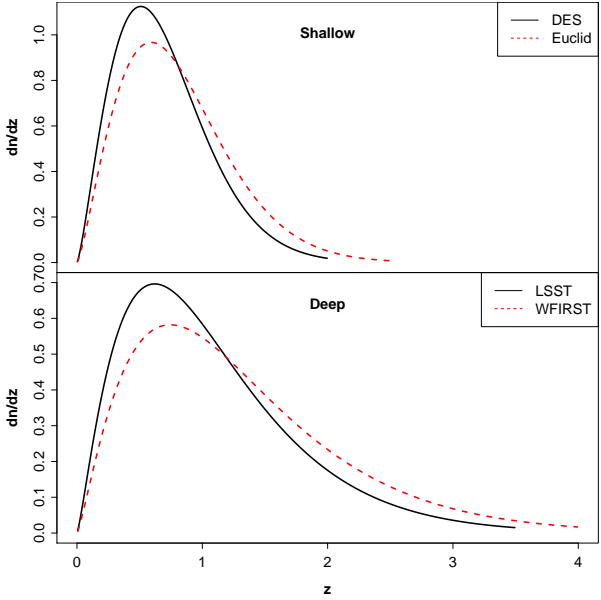}
\caption{Normalized redshift distributions for the four different surveys. \tbf{Top panel:} DES and Euclid. \tbf{Bottom panel:} LSST and WFIRST.}
         \label{fi:zdistrib}
\end{figure}

For our IA study we consider four different surveys: DES, Euclid, LSST and WFIRST. The corresponding redshift distributions are modeled as 
\be 
\label{redshiftben}
n(z)=N \,  z^\alpha \exp \left[ - \left(  \frac{z}{z_0} \right)^\beta \right]\,.
\ee 
The redshift distributions are shown in Fig. \ref{fi:zdistrib} with mean and median redshift summarized in Table \ref{tab:survey}. For LSST we adopt the redshift distribution suggested in \cite{cjj13}, and the DES redshift distribution is modeled by a modified CFHTLS redshift distribution \citep[see][adjusted for the slightly lower mean redshift of DES]{bhs07}. The Euclid redshift distribution follows a similar parameterization as DES, but is assumed to be slightly deeper. WFIRST redshifts are modeled after the LSST z-distribution, again assuming slightly deeper imaging. It is important to note that throughout this paper WFIRST refers to the wide-field imaging component of the WFIRST High-Latitude Survey (HLS) as described in \cite{sgb15}. The HLS component of WFIRST encompasses only 2 years and includes a substantial amount of spectroscopic grism observations. 

\begin{table}
\caption{Survey parameters (see text for details).}
\begin{center}
\begin{tabular}{|l|c c c c c c c|}
\hline
\hline
Survey & $\Omega_{\rm s}$ $[\mr{deg^2}]$ & $\sigma_\epsilon$ &  $n_\mr{gal}$ &$z_\mr{max}$ & $z_\mr{mean}$ & $z_\mr{med}$ &$m_{r,\mr{lim}}$\\
\hline
DES & $5,000 $ & $0.26$  & $10$ & 2.0 & 0.69 &0.62 & 24.0\\
LSST & $18,000$ & $0.26$ &$26$ & 3.5 & 1.07 & 0.93 & 27.0\\
Euclid & $15,000$ & $0.26$ &$20$ & 2.5 & 0.8 & 0.74 & 24.5\\
WFIRST & $2,200$ & $0.26$ &$45$ & 4.0 & 1.27 & 1.11 & 28.0\\
\hline
\hline
\end{tabular}
\end{center}
\label{tab:survey}
\end{table}

\subsection{Source galaxy distribution} 
\label{sec:fred}
\begin{table}
\caption{Fiducial luminosity function parameters.}
\begin{threeparttable}
\renewcommand*\TPTnoteLabel[1]{\parbox[b]{3em}{\hfill#1\,}}
\begin{tabular}{|c|c c c c c | c c |}
\hline
\hline
 &$\phi^*_{0} [(h/\rm{Mpc})^3]^\dagger$ & $M_{r}^{*}(0.1)^\dagger$ & $\alpha^\dagger$ & $P^\dagger$ & $Q^\dagger$& $P^{\rm{D}\ddagger}$ & $Q^{\rm{D}\ddagger}$ \\
\hline
all &$9.4\times10^{-3}$ & -20.70 &-1.23 & 1.8& 0.7&$-0.30^\mathsection$ &1.23 \\
red &$1.1\times10^{-2}$ &-20.34 &-0.57 &-1.2 & 1.8& $-1.15^\mathsection$&1.20 \\
\hline
\hline
\end{tabular}
\begin{tablenotes}
\small
\item $^\dagger$ $^{0.1}r$-band LF fit parameters from Tab.~5 in \citet{LFGAMA}.
\item $^\ddagger$ $B$-band LF evolution parameters from Tab.~6 in \citet{LFDEEP2}. 
\item $^\mathsection$ Value rescaled to match definition in Eq.~(\ref{eq:Pz}).
\end{tablenotes}
\end{threeparttable}
\label{tab:LF}
\end{table}
We use an evolving Schechter luminosity function (LF) $\phi(L,z)$ to model the LF of source galaxies
\be
\phi(L,z) = \phi^*(z) \left(\frac{L}{L^*(z)}\right)^\alpha \exp\left(-\frac{L}{L^*(z)}\right)\,,
\ee
with evolving Schechter parameters
\be
\phi^*(z) = \phi^*_{0} 10^{0.4Pz}\,,
\label{eq:Pz}
\ee
and $L^*(z)$ the luminosity corresponding to
\be
M^*(z) = M^*(0.1) - Q\left(z-0.1\right)\,.
\ee
For the baseline model this analysis uses the $r$-band fit parameters from the GAMA survey \citep{LFGAMA}, which are listed in Tab.~\ref{tab:LF}.

As described in Sect.~\ref{sec:Aia}, we follow \cite{jma11} in modeling the luminosity and redshift dependence of the IA amplitude. We evaluate the effective amplitude $\langle A \rangle$ and the fraction of contaminated (red) galaxies $f_\mr{red}$ from the LF of all and red galaxies
\bea
\nonumber A(m_\mr{lim},z)  &=&  A\left(L_0,z\right)\, \Bigg \langle \left(\frac{L}{L_0}\right)^\beta\Bigg \rangle_{\phi_\mr{red}} \\
&=& A\left(L_0,z\right) \frac{\int_{L(m_\mr{lim},z)}^\infty d L \left(\frac{L}{L_0}\right)^\beta\phi_\mr{red}(L,z)}{\int_{L(m_\mr{lim},z)}^\infty d L \phi_\mr{red}(L,z)}\label{eq:ampli}\,\\
f_\mr{red}(m_\mr{lim},z)& =&\frac{\int_{L(m_\mr{lim},z)}^\infty d L \phi_\mr{red}(L,z)}{\int_{L(m_\mr{lim},z)}^\infty d L \phi_{\rm{all}}(L,z)}\label{eq:fred}\,
\eea
with $L(m_\mr{lim},z)$ the $k$-corrected limiting luminosity at redshift $z$ in a survey with limiting apparent magnitude $m_\mr{lim}$ (c.f. Tab.~\ref{tab:survey} for values used in this analysis to model different surveys). $L(m_\mr{lim},z)$ is the luminosity corresponding to 
\be
M_{\rm{lim}}(z,m_{\rm{lim}}) = m_{\rm{lim}} - \left(5 \log_{10}\frac{D_{\rm L}(z)}{\rm{Mpc}/h}+25 +k(z)\right)\,,
\ee
where $D_{\rm L}(z)$ is the luminosity distance, and $k(z)$ the $k+e$-correction, for which we use the $r$-band values from \citep{Poggianti} due to the wide redshift range of this model. 

\begin{figure}
\includegraphics[width=8.5cm]{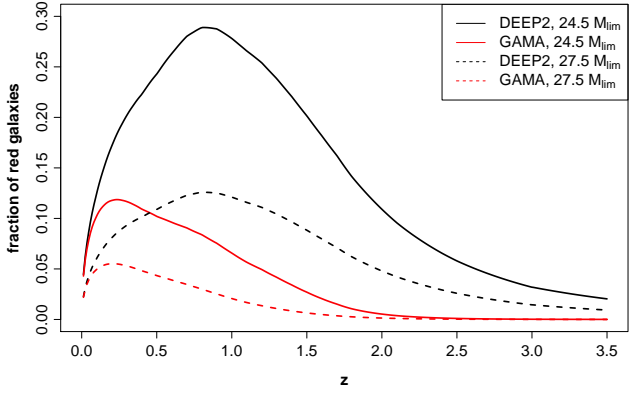}
\caption{Fraction of red galaxies computed from GAMA and DEEP2 luminosity function, respectively. We consider a deep and a shallow survey with limiting magnitude of 27.5 (LSST/WFIRST) and 24.5 (Euclid/DES), respectively.}
         \label{fi:fred}
\end{figure}

We note that the LF parameters and $k+e$-correction model assumed in this study differ from the choices in \citet{jma11}. While the \citet{LFDEEP2} LF parameter used in \citet{jma11} are based on deep observations LF measurements  to $z\sim 1$, compared to $z<0.5$ for GAMA, it is important to note that the former are $B$-band LF parameters (compared to $^{0.1}r$ for the GAMA parameters in our baseline model). Since Schechter parameters vary significantly across bands, it is not clear which of these will be a better model for the source galaxies of future weak lensing surveys with high sensitivity in the red and near-infrared bands. Hence the LF parameters are a major uncertainty in our forecasts for the unmitigated impact on IA on cosmic shear surveys. To illustrate the impact of LF parameters, we will consider a second LF model which combines the DEEP2 evolution parameters with the GAMA LF at the pivot redshift ($P^{\rm D}, \, Q^{\rm D}$ in Tab.~\ref{tab:LF}), while this differs from the \citet{LFDEEP2} LF we will refer to as DEEP2 LF in the following for simplicity. Figure \ref{fi:fred} shows $\fred$ derived from the  GAMA and DEEP2 LFs as a function of redshift. The large discrepancy between the curves indicates that the LF modeling of upcoming and future surveys limit is a major uncertainty for current forecasts. Compared to the uncertainty in LF evolution, the spread between $k+e$-correction templates is a minor source of uncertainty. 

\section{Results}
\label{sec:like}
\subsection{Simulated Likelihood analyses}
\label{sec:like_theory}
Throughout this study, we quantity the impact of IA on cosmological parameters using simulated likelihood analyses as described in this section. We generate a IA contaminated data vector as 
\be
\label{eq:datav}
\D = C^{ij} (l)+ C^{ij}_\mr{II} (l)+C^{ij}_\mr{GI} (l)
\ee
where the individual terms are calculated from Eqs. (\ref{eq:pdeltatopkappa}), (\ref{eq:IA1}), and (\ref{eq:IA2}). We split the source galaxies into five tomographic bins with equal number density of galaxies. From these five tomographic bins we derive five auto and ten cross power spectra, which we sample with 12 logarithmically spaced bins in $l$ ranging from $l_{\rm{min}} = 100$ to $l_{\rm{max}} = 5000$. 

The model vector is computed in a similar way; depending on whether we want to examine the impact of an IA scenario or mitigate IA, we calculate $\M = C^{ij} (l)$ or $\M = C^{ij} (l)+ C^{ij}_\mr{II} (l)+C^{ij}_\mr{GI} (l)$, respectively. While we contaminate the data vector with range of IA models (c.f. Sect.~\ref{sec:IAmod}), the mitigation strategy used throughout this analysis only uses the NLA Halofit IA model to compute the IA contribution to the model data vector $\M$.

Using the covariance defined in Sect. \ref{sec:covbasics} we compute the likelihood of the (IA contaminated) data vector $\D$ given a set of cosmological parameters $\pco$ as 
\bea
\label{eq:like}
 \like (\D| \pco) &=& N \, \times \, \exp \biggl[ -\frac{1}{2} \left( (\D -\M)^t \, \matC^{-1} \, (\D-\M) \right)  \biggr] \,,
\eea
where we have assumed that the errors of the data vector are distributed as a Multivariate Gaussian. $\matC$ is the covariance matrix, which we approximate to be independent of any of the parameters. We note that assuming a constant, known covariance matrix $\matC$ is an approximation to the correct approach of a cosmology dependent or estimated covariance \citep[see][for further details]{esh09}.

The posterior probability for a point in cosmological parameters $\pco$ is obtained via Bayes' theorem
\be
\label{eq:bayes}
\prob(\pco |\D) \propto \probr (\pco) \,\like (\D| \pco),
\ee
where $\probr (\pco)$ denotes the prior probability. 

As discussed in Sect.~\ref{sec:fred}, the luminosity function evolution parameters and the faint end slope of the luminosity function are key uncertainties for the amplitude of the IA contamination, and we treat these parameters as part of our IA model. In absence of reliable LF parameter constraint estimates for future surveys, we include abundance information to provide an implicit prior on LF parameters. Specifically, we include the number density of all and red galaxies at the mean redshift of each tomography bin, assuming that these will be measured with $5\%$ uncertainty. In addition to this abundance prior, we reject LF parameter combinations which cause unphysical models with $f_{\rm{red}} > 1$, which are an artifact of the phenomenological choice of parameterization.

We run simulated likelihood analyses in seven-dimensional parameter space (see Table \ref{tab:cosmology}) and integrate over ten IA nuisance parameters (see Table \ref{tab:nuisance}) when the CosmoLike IA mitigation module is used. This marginalized likelihood reads
\begin{align}
\label{eq:likemarg}
\nonumber L (\D|\pco ) =& \int \d \vek{ \pnu}  \, \exp \biggl( -\frac{1}{2} 
(\D - \M(\pco,\pnu))^t \matC^{-1} (\D - \M(\pco,\pnu)) 
\biggr)\\
& \times \exp\biggl[ -\frac{1}{2}\sum_{j\in\left\{all,red\right\}}\sum_{i=1}^5\frac{\left(\hat{ n}_j(z_i)-n_{j}( z_i,\vek{p})\right)^2}{\sigma^2(n_j(z_i))}\biggr] \,,
\end{align}
with $\pnu$ spanning the ten nuisance parameter dimensions, and where $\hat{n}_j(z_i)$ is fiducial galaxy abundance, $n_j(z_i,\vek{p})$ the galaxy abundance prediction for model parameters $\vek{p}$.

\subsection{LSST baseline models}
\label{sec:base}
\begin{table}
\caption{Fiducial cosmology, minimum and maximum of the flat prior on cosmological parameters.}
\begin{center}
\begin{tabular}{|l|c c c c c c c|}
\hline
\hline
&$\om$ & $\sig$ & $\ns$ & $\w$ & $\wa$ & $\omb$ & $h_0$\\
\hline
Fid & 0.315 & 0.829 & 0.9603 & -1.0 & 0.0 & 0.049 & 0.673\\
Min & 0.1 & 0.6 & 0.85 & -2.0 & -2.5 & 0.04 & 0.6\\
Max & 0.6 & 0.95 & 1.06 & 0.0 & 2.5 & 0.055 & 0.76\\
\hline
\hline
\end{tabular}
\end{center}
\label{tab:cosmology}
\end{table}

\begin{table*}
\caption{\tbf{Left:} Fiducial, minimum and maximum values (flat priors) for the intrinsic alignment parameters. \tbf{Right:} Fiducial values and range of the Gaussian priors for the nuisance parameters describing photo-z uncertainties (optimistic and pessimistic scenario) and uncertainties in the luminosity function. See text for details.}
\begin{tabular}{|l|c c c c|}
\hline
\hline
&$A_0$ & $\beta$& $\eta$& $\eta^\mr{z}$  \\
\hline
Fid  &5.92&1.1&-0.47 & 0.0\\
Min  &0.0 &-5.0 &-10.0 &-3.0 \\
Max &10.0 &5.0&10.0 &3.0 \\
\hline
\hline
\end{tabular}
\hspace{0.5cm}
\begin{tabular}{|l|c c c c c c c c|}
\hline
\hline
&$\mr{bias}_\mr{ph}$&$\sigma_\mr{ph}$& $ \Delta \mr{LF}_\alpha$ & $\Delta \mr{LF}_\mr{P}$ & $\Delta \mr{LF}_\mr{Q}$ & $\Delta \mr{LF}_\alpha^\mr{red}$ & $\Delta \mr{LF}_\mr{P}^\mr{red}$& $\Delta \mr{LF}_\mr{Q}^\mr{red}$ \\
\hline
Fid & 0.0 &0.05  &0.0&0.0&0.0&0.0 &0.0&0.0\\
$\sigma$ (Gaussian Prior) &0.002  &0.003 & 0.05&0.5& 0.5&0.1 & 0.5 & 0.5 \\
$\sigma$ (Gaussian Prior) &0.005  &0.006 &  0.05&0.5& 0.5&0.1 & 0.5 & 0.5\\
\hline
\hline
\end{tabular}
\label{tab:nuisance}
\end{table*}

\begin{figure*}
\includegraphics[width=17cm]{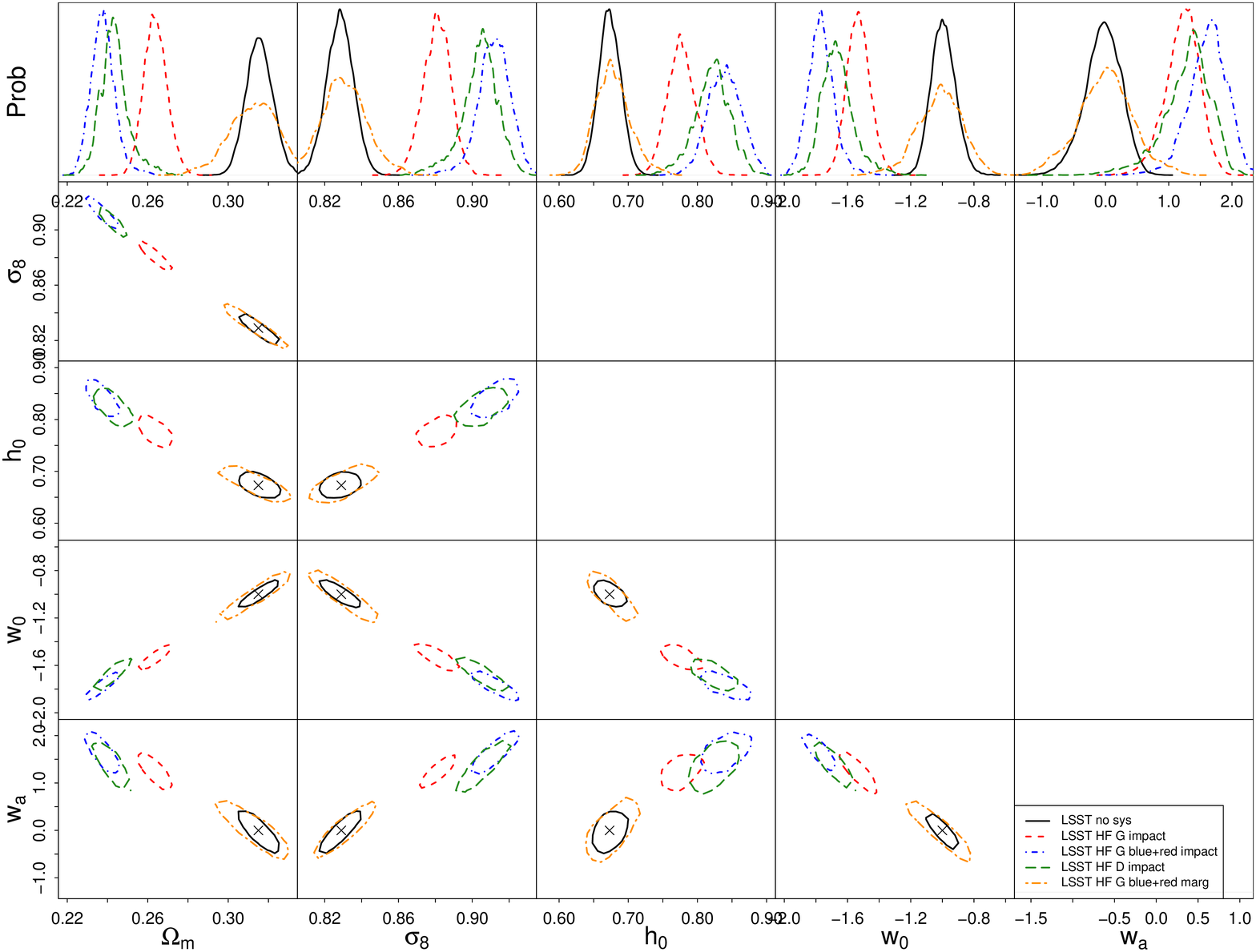}
\caption{The impact of IA on WL constraints ($68\%$ confidence region) from LSST assuming the NLA Halofit scenario. We consider different luminosity functions i.e. GAMA (red/dashed) and DEEP2 (green/long-dashed) and for the GAMA LF we also consider the case for which blue galaxies have a mild NLA IA contribution (blue/dot-dashed). The LSST statistical errors are shown in black/solid. Orange/dot-long-dashed contours show results when using the most extreme of these cases, i.e. the data vector corresponding to the blue contours, as input and including the CosmoLike IA mitigation module in the analysis. The marginalized likelihood is obtained by integrating over a 11-dimensional nuisance parameter space (see text for details).}
         \label{fi:lsst1}
\end{figure*}

\begin{figure*}
\includegraphics[width=17cm]{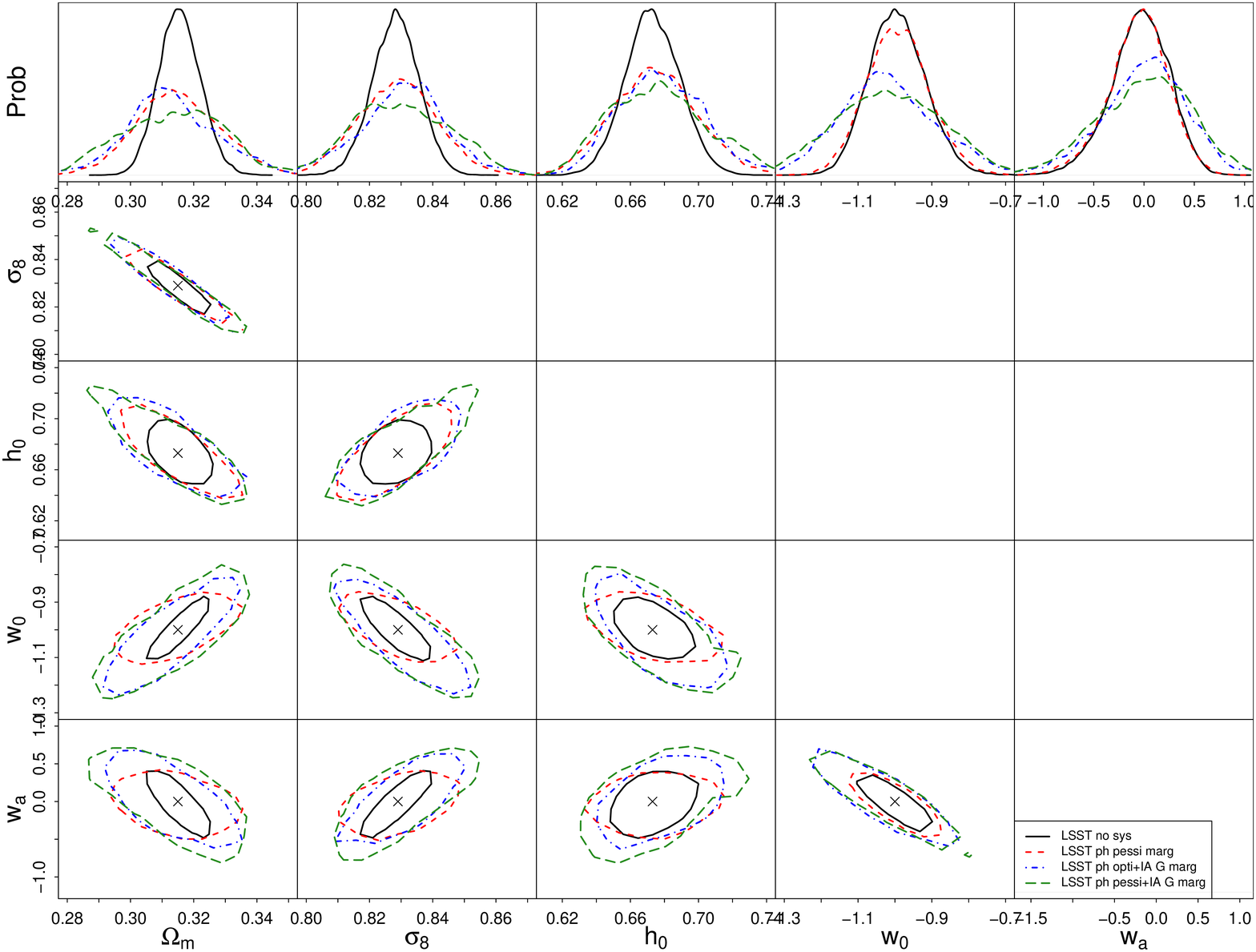}
\includegraphics[width=17cm]{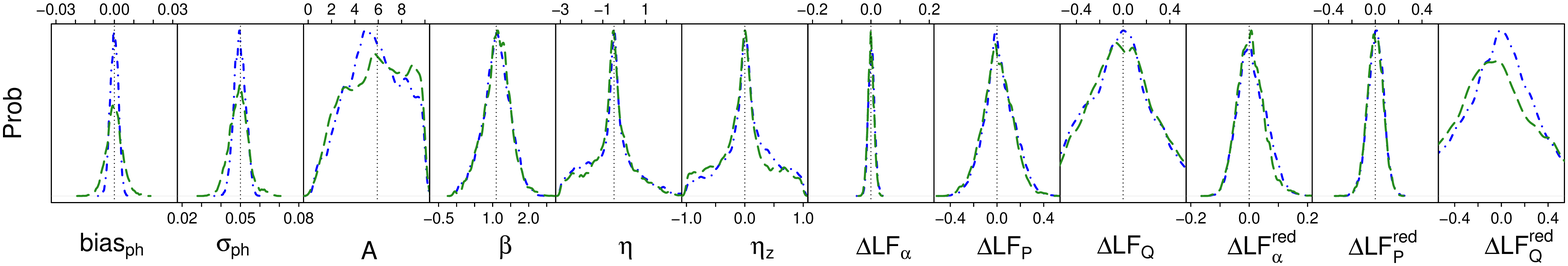}
\caption{\tbf{Top}: Marginalized WL constrains ($68\%$ confidence region) from LSST when marginalizing over Gaussian photo-z uncertainties (red/dashed) and joint uncertainties of photo-z's and the fiducial IA NLA Halofit model. We assume two different levels of photo-z errors resembling optimistic and pessimistic LSST photo-z errors. The black/solid lines again show the LSST statistical errors for comparison. \tbf{Bottom:} The marginalized one-dimensional posterior probabilities for the 12 nuisance parameters used in the optimistic and pessimistic joint photo-z and IA analysis.}
         \label{fi:lsst2}
\end{figure*}

In this section we present detailed likelihood analyses for LSST assuming the NLA Halofit scenario.
\paragraph*{Impact of luminosity functions}
As detailed in Sect. \ref{sec:fred}, the extrapolation of existing LF measurements to the depth of future cosmic shear surveys is a major uncertainty in modeling IA. In Fig. \ref{fi:lsst1} we illustrate the range of this uncertainty by running simulated likelihood analysis using the GAMA \ti{(red/dashed)} and DEEP2 \ti{(green/long-dashed)} luminosity functions in the IA contaminated data vector. Both analyses assume the NLA Halofit scenario in the data vector; the corresponding contours mimic likelihood analyses which do not account for the IA contamination at all. As a result, corresponding constraints are biased at the level of several $\sigma$ (the black ``x" indicates the input cosmology). The LSST statistical errors are shown in \ti{black/solid}. We also run likelihood analyses that include the CosmoLike IA mitigation module, but defer the discussion of these results to Sect. \ref{sec:comp}. 
 
\paragraph*{Alignment of blue galaxies}
While measurements of the position--shape correlation for blue galaxies indicate that the NLA signal of blue galaxies is consistent with zero \citep[e.g.,][]{hmi07,mbb11}, the statistical uncertainty in these non-detections is substantial and does not rule out noticeable contamination. Hence in Fig. \ref{fi:lsst1} we also illustrate the impact of NLA contamination for blue galaxies with an amplitude based on the $68\%$ upper limit for the NLA amplitude of WiggleZ and blue, low-redshift SDSS galaxies in \citet{mbb11}. Assuming the same redshift and luminosity scaling of the blue galaxy IA amplitude as in our fiducial model for red galaxies, in our notation this corresponds to an amplitude $A(L_0, z)_\mr{blue}=0.25 \times A(L_0, z)_\mr{red}$. 
The \ti{blue/dot-dashed} contours in Fig. \ref{fi:lsst1} correspond to a scenario where both red and blue galaxies are affected by IA, and the total IA amplitude is calculated by 
\be
A_{\rm{tot}}(m_{\rm{lim}},z) = f_{\rm{red}}A_{\rm{red}}(m_{\rm{lim}},z)  + (1-f_{\rm{red}})A_{\rm{blue}}(m_{\rm{lim}},z)\,.
\ee 
This analysis again assumes the NLA Halofit scenario with the GAMA luminosity function, the IA model for red galaxies described in Sect. \ref{sec:iabasics}, and the blue galaxies NLA model motivated above. Equation \ref{eq:ampli} for blue galaxies hence uses $A(L_0, z)_\mr{blue}=0.25 \times A(L_0, z)_\mr{red}$, and we approximate $\phi(L,Z)_\mr{blue}$ with $\phi(L,Z)_\mr{all}$. We note that a more realistic IA model for blue galaxies involves will include a contribution from the tidal torquing mechanism; we defer a detailed analysis to future work (Zu et al., in prep.).

The \ti{orange/dot-long-dashed} contours in Fig.~\ref{fi:lsst1} show results when marginalizing over the IA nuisance parameters as described in Eq. \ref{eq:likemarg}. In addition to the four IA and six luminosity function nuisance parameters (see Table \ref{tab:nuisance}), in this particular analysis we include an additional nuisance parameter describing the fractional amplitude of the blue IA signal (centered around the fiducial value of 0.25). 
The bias from IA contamination (\ti{blue/dot-dashed}) is completely removed in the case of the marginalized likelihood analysis; the broadening of the likelihood contours is noticeable but not excessive \citep[c.f.][for the degradation of other IA mitigation techniques]{kirk12,kbh15}.
\paragraph*{Impact of photometric redshift uncertainties}
Figure \ref{fi:lsst2} shows the increase in error bars when marginalizing over both IA and photo-z uncertainties. We model the latter assuming that a Gaussian photo-z probability distribution function is obtained from each galaxy. Uncertainties in the overall redshift distributions of each tomographic bin are then parameterized as variations in the mean (``bias") and variations in the width of said Gaussian ``$\sigma$". We further assume Gaussian priors on the bias and $\sigma$ parameters, i.e. we consider an optimistic photo-z scenario with $\Delta \mr{bias}=0.002$, $\Delta \mr{\sigma}=0.003$, and a pessimistic scenario with $\Delta \mr{bias}=0.005$, $\Delta \mr{\sigma}=0.006$. 
 
The red/dashed contours in Fig. \ref{fi:lsst2} show results when marginalizing over photo-z uncertainties only. We see a substantial degradation of constraints on $\om$, $\sig$, and to a lesser extent on $w_0$. Accounting for IA and photo-z jointly also substantially degrades $w_a$ and $n_s$. In the presence of IA the two levels of photo-z errors we consider hardly change the contour size. When looking at the nuisance parameter panel of Fig. \ref{fi:lsst2} we see that although the photo-z parameters ($\mr{bias}_\mr{ph}$, and $\sigma_\mr{ph}$) are obviously better constrained in the optimistic photo-z prior scenario, the additional information gets largely absorbed in the luminosity function parameters. 

We emphasize that the Gaussian photo-z model is optimistic in the sense that it neglects catastrophic outliers, which have been shown to severely degrade dark energy constraints \citep{beh10}. We postpone a more quantitive analysis of the degeneracies of IA, photo-z, and cosmological parameters to future work. 

\paragraph*{Varying the IA model}
\begin{figure*}
\includegraphics[width=17cm]{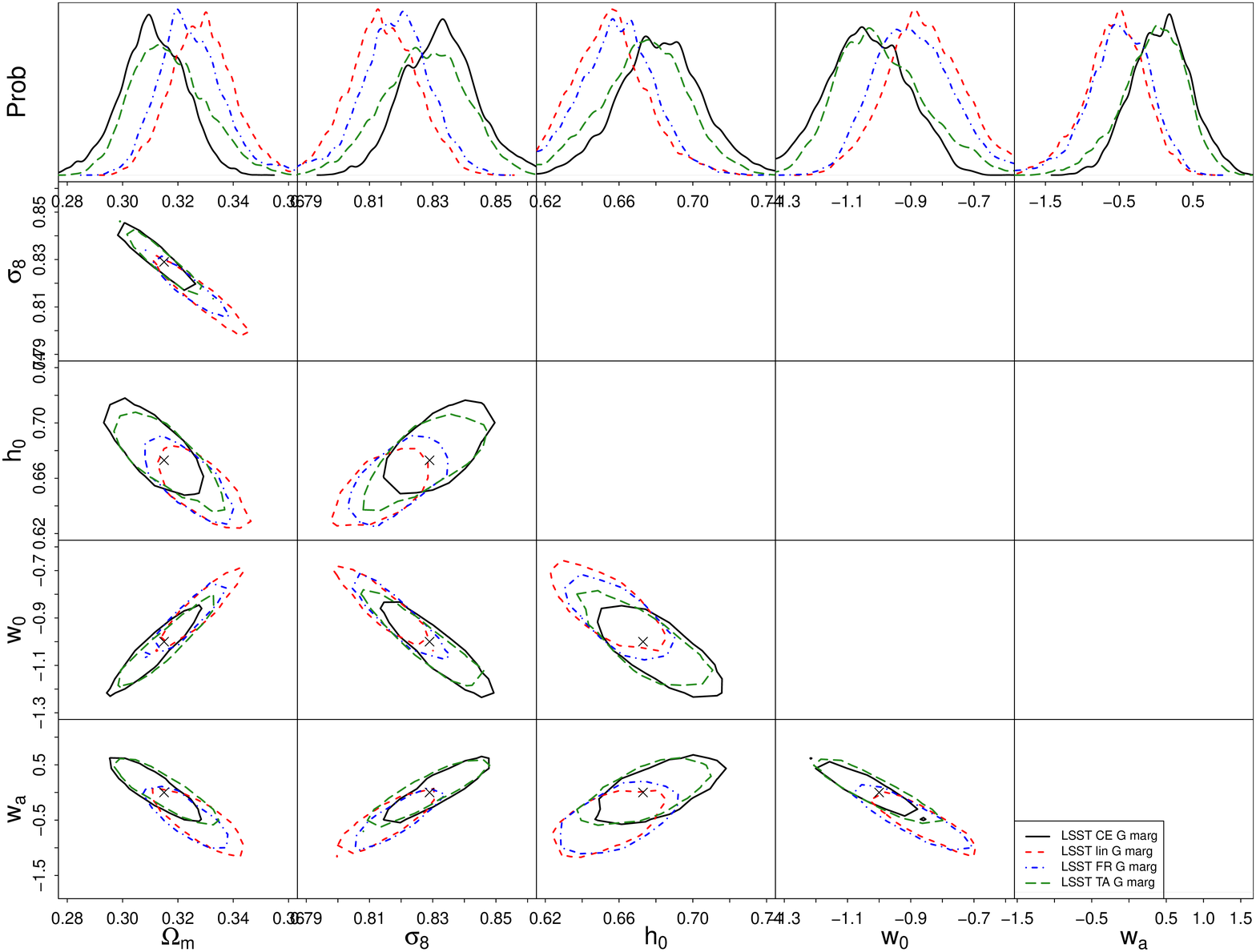}
\includegraphics[width=17cm]{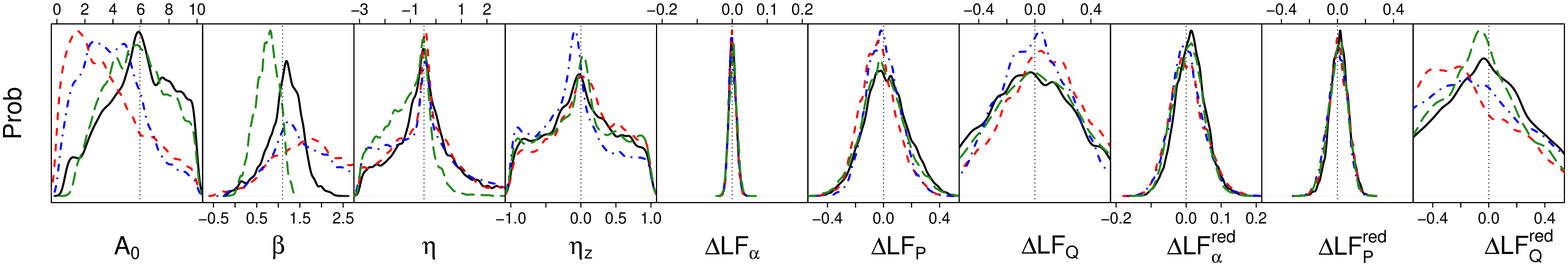}
\caption{\tbf{Top}: WL constraints ($68\%$ confidence region) from LSST when assuming different models in the data vector, but assuming the fiducial IA NLA Halofit model in the marginalization. The black/solid lines show results from a data vector contaminated by the NLA Coyote Universe model, which is very close to the fiducial NLA Halofit scenario. Red/dashed corresponds to the linear power spectrum model, blue/dot-dashed to the freeze-in model, green/long-dashed to the tidal alignment model. All models assume the GAMA luminosity function. \tbf{Bottom:} The marginalized one-dimensional posterior probabilities for the ten nuisance parameters describing the IA uncertainty.}
         \label{fi:lsst3}
\end{figure*}

We now extend our analysis to all IA scenarios described in Sect. \ref{sec:IAmod}. Specifically, we create IA contaminated data vectors for the NLA Cosmic emulator, the Freeze-in, the linear, and the tidal alignment model, and then analyze these data vectors assuming the NLA Halofit model in the marginalization. This analysis mimics the realistic scenario where the analyst has a broad idea, but not an exact model of the IA contamination in the data.

Figure \ref{fi:lsst3} shows the 68\% likelihood contours for the aforementioned four IA scenarios. The NLA Cosmic Emulator (black/solid) and the tidal alignment (green/long-dashed) scenarios are well captured by the NLA Halofit model. The linear power spectrum scenario (red/dashed) shows the largest offset from the underlying fiducial model, the freeze-in scenario's offset (blue/dot-dashed) is smaller but noticeable. Even for the worst case, i.e. the linear IA contamination mitigated using a NLA model, the fiducial cosmological model is within the 68\% likelihood contours. As our likelihood analysis assumes $l_\mr{max}=5000$, where differences between linear and non-linear power spectra are significant, we conclude that when IA is fully described by some tidal alignment scenario, an exact IA model is not vital for removing biases in cosmological parameters. We will consider different surveys and the impact of the galaxy luminosity function in the following subsection.

\subsection{Comparison of LSST, DES, Euclid, WFIRST}
\label{sec:comp}
In the previous sections we examined the impact of IA contamination on LSST WL constraints; we now extend our analysis to WFIRST, Euclid, and DES. For the comparison of these surveys it is important to note that the significance of IA biases, and hence the importance of IA as a contaminant, is affected by survey depth and survey width: The amplitude of the IA contamination relative to the cosmic shear signal is highest at low redshift and for more luminous source galaxies, hence the amplitude of the parameter bias decreases with survey depth. The significance of the contamination is given by the ratio of the amplitude of the parameter bias and the survey's constraining power on these parameters, which increases with survey depth and width.

Figure \ref{fi:1d1} shows the marginalized one-dimensional $68\%$ error bars on cosmological parameters for the deep surveys LSST ($m_r =27$) and WFIRST ($m_r =28$). Figure \ref{fi:1d2} shows the same analysis for the shallower surveys Euclid ($m_r = 24.5$) and DES ($m_r = 24$). Each plot comprises results of 22 different likelihood analyses; the presentation of the results for different surveys follow the same structure: 

\begin{figure*}
\includegraphics[width=17.0cm]{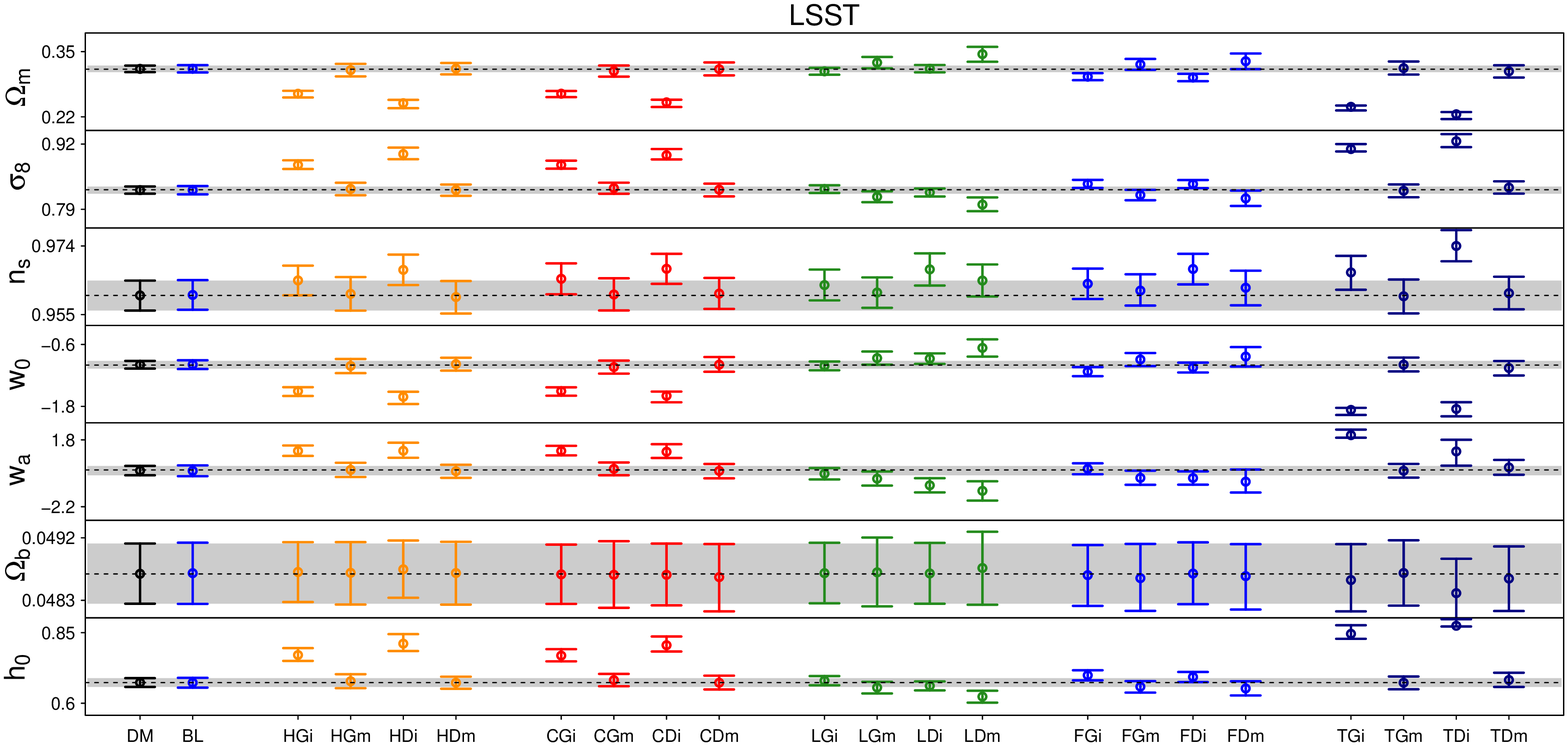}
\includegraphics[width=17.0cm]{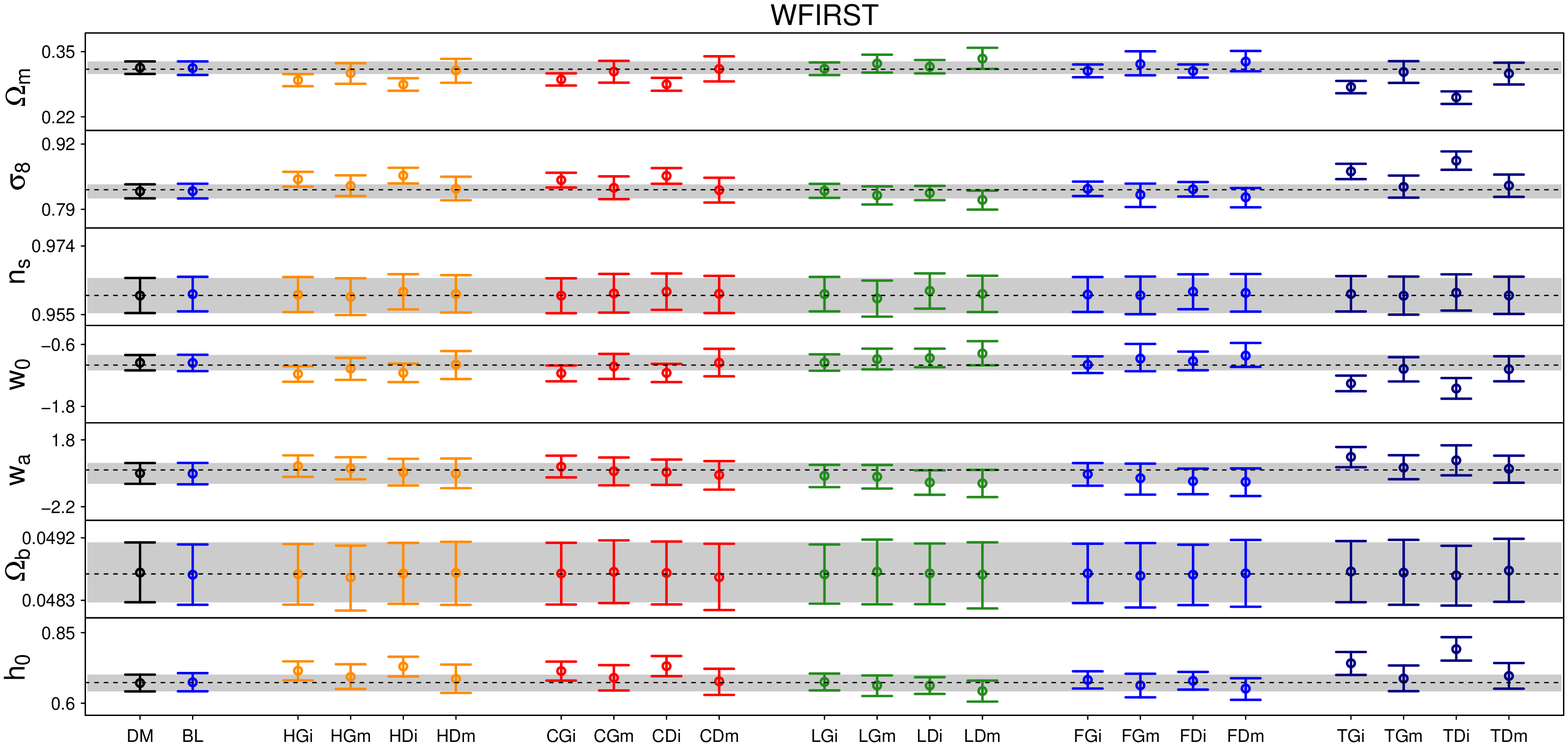}

\caption{One-dimensional 68\% error bars on cosmological parameters for the deep surveys LSST and WFIRST. The abbreviations that reference the different models read ``DM'' :=Dark matter, ``BL'' := DM with Blue Galaxies only, ``H'' := NLA Halofit, ``C'':=NLA Coyote Universe, ``L'':= LA linear power spectrum, ``F'':= freeze-in, ``T'':= tidal alignment. For each model we run four different analyses, considering impact ("i") of untreated IA contamination and marginalized (``m'') results using the NLA Halofit model to account for the IA uncertainty. For all IA scenarios we consider both GAMA and DEEP2 luminosity functions, denoted as ``G'' and ``D'', respectively. }
         \label{fi:1d1}
\end{figure*}

\begin{figure*}
\includegraphics[width=17.0cm]{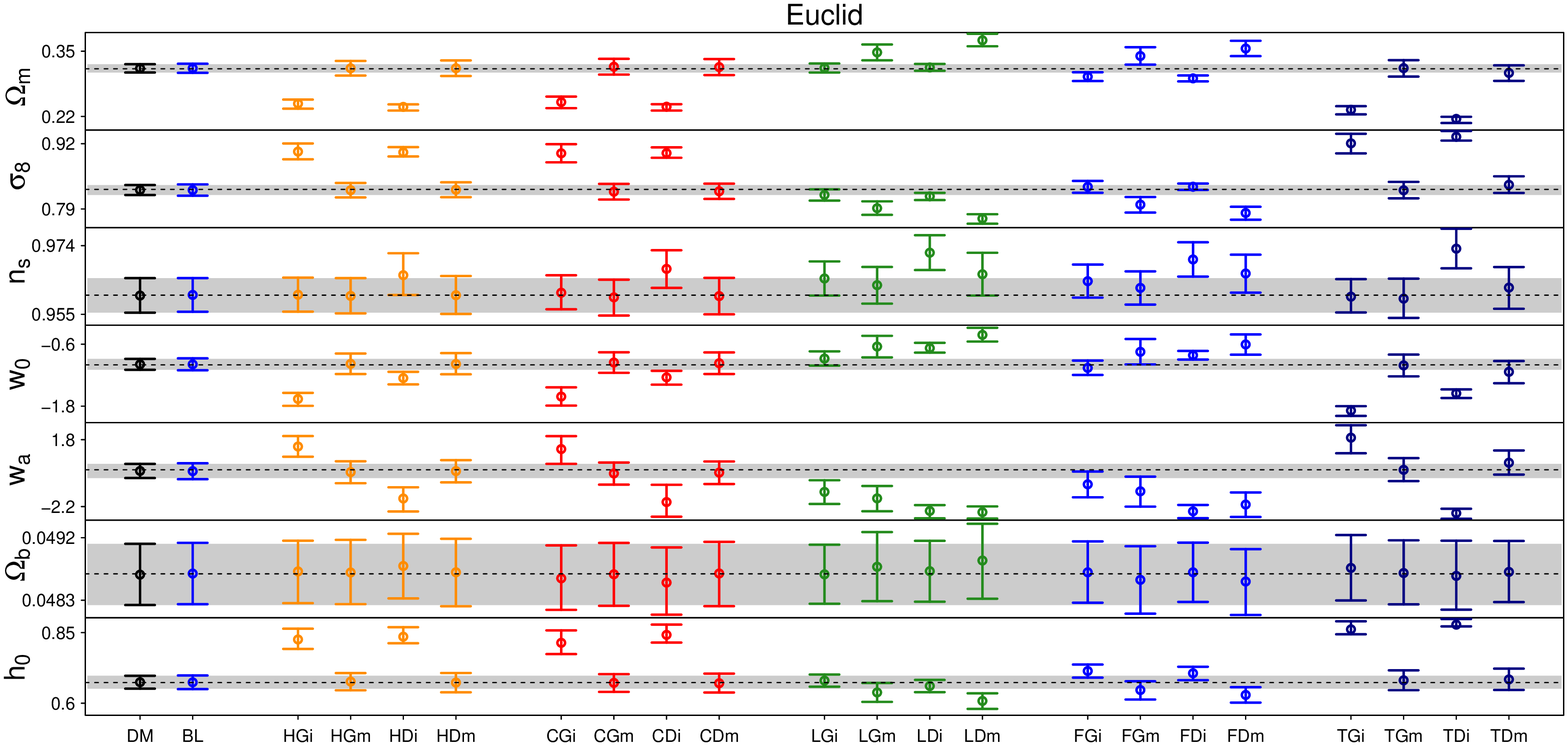}
\includegraphics[width=17.0cm]{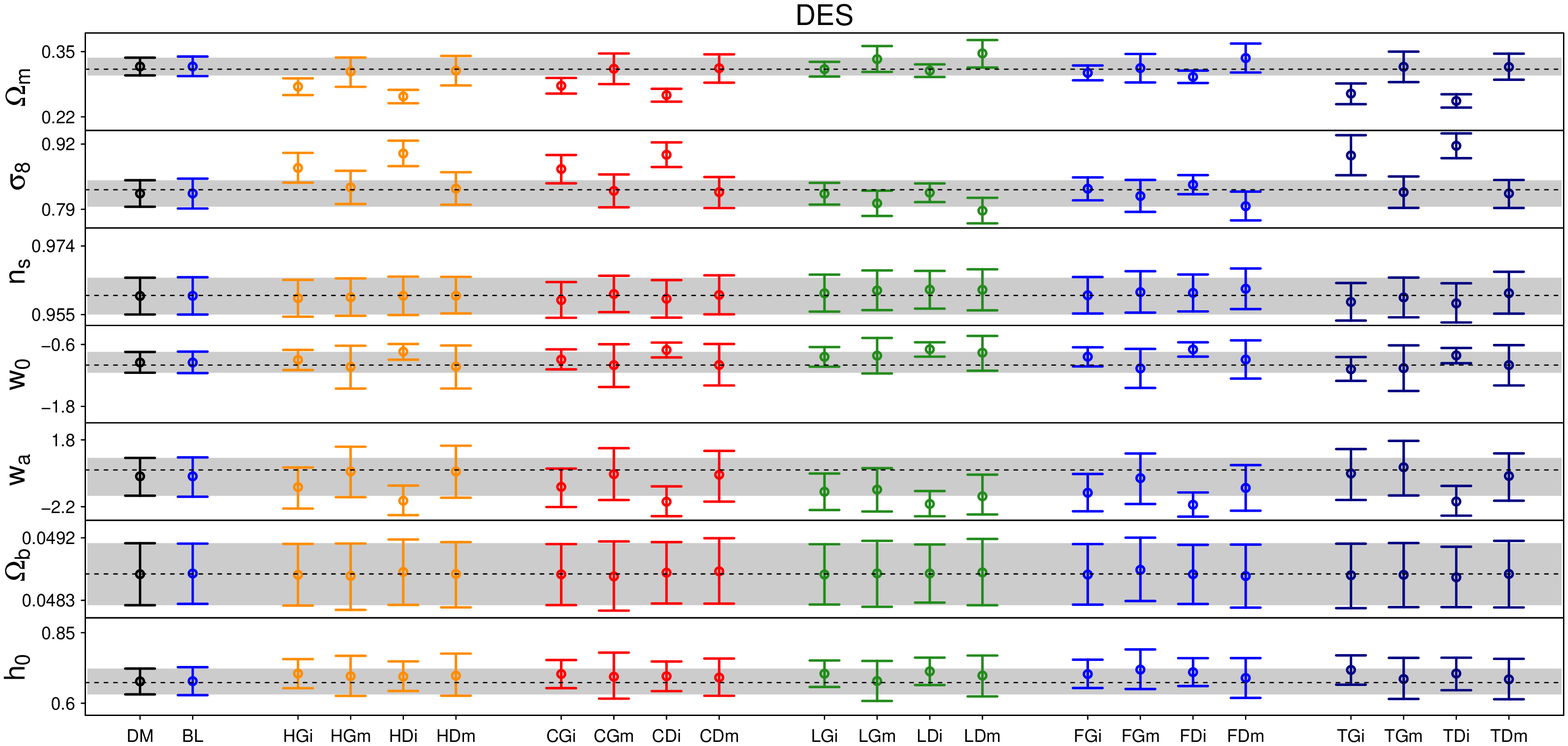}

\caption{Same as Fig.~\ref{fi:1d1}, but for the shallow surveys DES and Euclid.}
         \label{fi:1d2}
\end{figure*}

We first show the statistical error bars (black) obtained from using an uncontaminated DM data vector as an input and switching off the CosmoLike IA module in the simulated likelihood analysis. In blue we show a similar analysis but reducing the number density of source galaxies by 20\% (denoted as ``BL"), which is a simplified implementation of identifying the severely IA contaminated (red) fraction of galaxies and excluding these galaxies from the analysis (c.f. Fig.~\ref{fi:fred} for the expected red fraction of different surveys).

We note that the reduction in source density only causes a minor loss of cosmological constraining power (also see Fig. \ref{fi:Euclid_alter}). While this result might be counter-intuitive given the apparent importance of shape noise in many forecasts, it can be explained by the use of non-Gaussain covariance matrices. For Gaussian covariances only the cosmic variance term is independent of $n_\mr{gal}$, whereas a larger $n_\mr{gal}$ substantially reduces the noise and mixed term, which again substantially increases the information on small physical scales. Non-Gaussian covariances are dominated by terms that stem from higher-order moments of the density field and especially the Halo Sample Variance term, all of which are unaffected by $n_\mr{gal}$. This result emphasizes the prospect of selecting well-understood and characterized galaxy subsamples for the analysis, thereby avoiding potentially severe systematics while minimally degrading cosmological constraints.

After establishing the statistical error budget as a baseline we consider five IA scenarios (see Sect. \ref{sec:IAmod} for details) to contaminate the data vector. In order of their appearance in Figs. \ref{fi:1d1} and \ref{fi:1d2}  the considered models are: 
\begin{itemize}
\item Nonlinear linear alignment using Halofit $P_\delta$ (orange)
\item Nonlinear linear alignment using Coyote Universe $P_\delta$ (red)
\item Linear alignment using linear power spectrum \citep[transfer function from][]{eih98} (green)
\item Freeze-in model using the geometric mean of linear power spectrum and Halofit $P_\delta$ (blue) 
\item Full tidal alignment model using the Halofit $P_\delta$ (dark blue)
\end{itemize}

For each model we run four different analyses, two for the GAMA LF and two for the DEEP2 LF. Of these two the first analysis always shows the impact of the IA contaminated data vector in the absence of any mitigation and the second analysis shows constraints when using the NLA Halofit model in the CosmoLike IA mitigation module. The marginalized likelihood is obtained by integrating over a ten-dimensional IA nuisance parameter space as described in Eq. (\ref{eq:likemarg}). 

In addition to Figs. \ref{fi:1d1} and \ref{fi:1d2}, we evaluate the magnitude of the bias in cosmological parameters from all likelihood analyses using a similar metric as in \cite{ekd14}. The bias is defined as the difference between the best-fit value of the parameter emerging from the likelihood analysis and the fiducial value used to generate the spectra. We then compare this bias to the size of the seven-dimensional cosmological likelihood volume at different confidence levels and asses whether our IA mitigation strategy is successful. 

If we considered only one cosmological parameter the relevant quantity that characterizes bias with respect to likelihood volume would be $(p^{\rm best\, fit} - p^{\rm fid})^2/\sigma^2$. 
For our seven parameter case, this concept generalizes to
\be
\label{eq:paradistance}
\Delta \chi^2 = (\pco^\mr{fid} - \pco^\mr{IA, best \, fit})^t \, \matC_{\pco}^{-1} \, (\pco^\mr{fid} - \pco^\mr{IA, best\, fit})
,\ee
where the covariance matrix is determined via
\be
\label{eq:paracov}
\matC_{\pco}^{ij} = \frac{1}{N-1} \sum_{k=0}^N \left(\langle \pco^i \rangle - \pco^{ik} \right) \left(\langle \pco^j \rangle - \pco^{jk}\right)\,,
\ee
with $\langle \pco^i \rangle$ indicating the mean of the $i$-th cosmological parameter ($i,j \in [1,7]$), and $k \in [1,N]$ being the index running over all steps in the MCMC chain. Assuming a $\chi^2$ distribution with seven degrees of freedom, we find the critical $\Delta \chi^2$ values that correspond to $68\%$, $95\%$, and $99\%$ confidence regions are 8.14, 14.07, and 18.48, respectively. We summarize corresponding values in Table \ref{tab:paradist} and find a qualitative similar behavior as in the Figs. \ref{fi:1d1} and \ref{fi:1d2}. 

It is important to note that $\Delta \chi^2$ and the metric employed in Figs. \ref{fi:1d1} and \ref{fi:1d2} quantify the success of IA mitigation differently. The latter accounts for the non-Gaussianity of the posterior distribution in parameter space, but projects the posterior probability onto a specific basis (i.e. the cosmological parameters). It is fair to assume that if all projections are free of residual biases, the overall bias for the seven-dimensional cosmological parameter space is under control. The $\Delta \chi^2$ values in Table \ref{tab:paradist} give an overall estimate of the residual bias in combination with the multi-dimensional error in parameter space, however under the explicit approximation that the posterior probability is distributed as a multivariate Gaussian.
\paragraph*{IA impact} For a given survey, the non-linear IA models (NLA Halofit, NLA Coyote, TA) in combination with the DEEP2 luminosity function give rise to the strongest IA contamination, and hence cause the largest biases in cosmological parameters when no mitigation is applied.

As a consequence of small statistical uncertainties and relatively shallow observations, Euclid is most severely affected by IA. Corresponding $\Delta \chi^2$ values in Table \ref{tab:paradist} range from 111 for the NLA Halofit to 91 for the TA scenario.  While LSST has similar constraining power as Euclid, the relative contribution of the IA contamination to the shear signal is reduced due to the deeper redshift distribution, and the corresponding parameter biases are smaller. For LSST the $\Delta \chi^2$ values for these IA scenarios range from 71 to 57. 

DES has substantially less constraining power than either Euclid or LSST, but it is also significantly shallower than LSST, and even slightly shallower than Euclid; as a consequence, despite the larger error bars, parameter biases due to unmitigated IA are still considerable and $\Delta \chi^2$ values for the non-linear IA models range from 58 to 54. The HLS component of the WFIRST survey with an area of 2200 deg$^2$ has higher constraining power than DES, but less than Euclid and LSST; due to the depth of the survey, $\Delta \chi^2$ values for these three IA scenarios range from 37 to16, which still constitutes a significant bias that must be addressed. 

Compared to the non-linear IA models, the IA contamination from the linear alignment and freeze-in model is much smaller (at large $l$), and the parameter biases are less severe.
For WFIRST these biases are below the 1-$\sigma$ threshold before any mitigation is applied, i.e.  $\Delta \chi^2 = 4.1$ and $\Delta \chi^2 = 2.3$ for freeze-in and linear alignment, respectively. 

\paragraph*{IA mitigation} The CosmoLike IA mitigation module (NLA Halofit model with 10 nuisance parameters, c.f. Sect.~\ref{sec:like_theory}) successfully removes IA biases for the NLA Halofit, NLA Coyote, and the TA scenarios. While this is expected for the NLA Halofit scenario as input (recall that we use this exact scenario in our mitigation module), it is encouraging to see that the technique is sufficiently robust to mitigate the other two scenarios with different prescriptions for the non-linear regime as well.  In particular, biases caused by the full tidal alignment model, which as \cite{bvs15} have shown is in excellent agreement with observations, are fully removed by our mitigation scheme. The corresponding $\Delta \chi^2$ values for this model assuming the DEEP2 LF and Euclid survey drop from 91 to 1.1 after mitigation, indicating that after IA mitigation the underlying cosmology can be recovered without significant biases.

For the linear alignment and freeze-in models mitigation is not as straightforward:
These models predict angular IA power spectra which differ substantially in terms of shape in $l$ and redshift scaling compared to the NLA Halofit template assumed in our mitigation strategy. As a consequence the success of the IA mitigation depends on how well the nuisance parameters can compensate these differences. In particular, if differences between these IA models are more degenerate with cosmological parameters than with nuisance parameters the IA mitigation will not be successful.

For the freeze-in model IA mitigation using the NLA Halofit works sufficiently well for the LSST, WFIRST and DES survey parameters. For Euclid however, the residual bias may be severe depending on the underlying luminosity function: we find $\Delta \chi^2 = 19$ and $\Delta \chi^2 = 7.4$ for the DEEP2 LF and GAMA LF parameters respectively. This indicates that the simple mitigation template proposed in the analysis may not be sufficient for Euclid if the IA contamination were similar to the freeze-in model. This is clearly a consequence of the fact that we use use the NLA Halofit model in our IA mitigation strategy independent of which scenario was assumed in the data vector. Even though the freeze-in model is based on a combination of linear and nonlinear power spectrum, its angular shape and redshift dependence are sufficiently different from the NLA Halofit that ten (amplitude related) nuisance parameters cannot compensate these differences. 

This problem becomes more prominent for linear alignment model as the underlying IA scenario. For Euclid, the residual bias after IA mitigation is significant for the GAMA LF model, and even more severe for the DEEP2 LF (corresponding values are $\Delta \chi^2 =11$ and $\Delta \chi^2 =29$, respectively). Even for LSST, the DEEP2 LF case after IA mitigation ($\Delta \chi^2 =11$) now exceeds the 1-$\sigma$ threshold ($\Delta \chi^2 =8.14$). 
The difference between linear and NLA IA contamination, and thus the failure of our mitigation strategy, is aggravated by the optimistic choice of scale $l_\mr{max}=5000$. For both, Euclid and LSST, we run additional chains with a reduced $l_\mr{max}=2000$; the corresponding $\Delta \chi^2$ values drop to 13 (from 29) and to 6 (from 11) for Euclid and LSST, respectively. This allows us to claim moderate success for LSST, Euclid however is still severely affected. 

For WFIRST, the proposed IA mitigation is successful; there was no large bias to begin with and mitigation shows a slight improvement, primarily due to a slightly increased error bar after marginalizing over the nuisance parameters. A similar statement may hold for DES, however the bias in case of the DEEP2 LF approaches the 1-$\sigma$ threshold. We discuss the implications of these findings on IA mitigation strategies for future surveys in Sect.~\ref{sec:disc}. 

In addition to the bias removal we consider the information loss as a major criteria for a successful mitigation scheme. We quantify this information loss ``IL" as the ratio of geometric mean computed from the 1-D cosmological parameter error bars of the considered scenario and the geometric mean statistical error. 
\be
\mr{IL}=\left(\frac{\prod_i \sigma^\mr{Scenario}_i} {\prod_i \sigma^\mr{DM}_i} \right)^{1/7}
\ee
Table \ref{tab:infoloss} shows the corresponding increase in error bars for five different scenarios, i.e. when removing 20\% of the galaxies, the NLA Coyote scenario, and the full tidal alignment scenario. The loss in constraining power when removing one fifth of the galaxies is minimal across all surveys; it ranges from 4-10\%. These numbers are higher when mitigating IA with 10 nuisance parameters and priors as described in Tab.~\ref{tab:nuisance}; the increase in error bars ranges from approximately 30\% to 55\% depending on survey and luminosity function. 

For the interpretation of these numbers it is important to note that  this analysis employed broader priors range than current observational limits for the NLA amplitude \citep[e.g.][]{jma11} and LF parameters, in order to enable the comparison of different IA models, such as the mitigation on linear alignments with a halofit NLA template, which requires normalization parameters outside the observed prior range for the NLA model. Hence the losses in constraining power reported in Table~\ref{tab:infoloss} are conservative estimates, the use of observed priors for the NLA amplitude and LF measurements from the data set will improve the performance of this mitigation strategy considerably.

\begin{table}
\caption{The $\Delta \chi^2$ distance (see Eq. \ref{eq:paradistance}) between best fit and fiducial parameter point. Cases where IA mitigation fails are indicated as red.}
\begin{center}
\def\arraystretch{1.3}
\begin{tabular}{|l| l l l l |}
\hline
Scenario & DES & Euclid & WFIRST & LSST   \\ \hline
DM&0.39&0.15&0.35&0.094  \\ 
BL&0.34&0.21&0.38&0.16  \\ \hline
HGi&22&65&6.6&61  \\ 
HGm&0.56&0.36&0.61&0.42  \\ 
HDi&55&111&18&73  \\ 
HDm&0.59&0.27&0.75&0.29  \\ \hline
CGi&22&63&5.9&55  \\ 
CGm&0.82&0.44&0.79&0.36  \\ 
CDi&58&99&16&71  \\ 
CDm&0.54&0.42&0.73&0.42  \\ \hline
LGi&3.4&15&0.94&5.8  \\ 
LGm&2.1& \textcolor{red}{11}&0.55&2.5  \\ 
LDi&11&33  &2.3&15  \\ 
LDi  ($l_\mr{max}=2000$)& - &39 & - &19  \\ 
LDm&5.9&\textcolor{red}{29} &1.4&\textcolor{red}{11}   \\
LDm ($l_\mr{max}=2000$) & - & \textcolor{red}{13} & - &6  \\ \hline
FGi&5.8&32&1.4&14  \\ 
FGm&1.8&7.4&0.82&1.8  \\ 
FDi&19&68&4.1&33  \\ 
FDm&3.3&\textcolor{red}{19}&0.91&3.8  \\ \hline
TGi&30&87&13&64  \\ 
TGm&1.5&0.8&0.55&0.69  \\ 
TDi&54&91&37&57  \\ 
TDm&1&1.1&0.99&0.77  \\ \hline
\end{tabular}
\end{center}
\label{tab:paradist}
\end{table}

\begin{table}
\caption{The loss in constraining power with respect to the statistical errors for five scenarios (see text for details). }
\begin{center}
\def\arraystretch{1.3}
\begin{tabular}{|l| l l l l |}
\hline
Scenario  & DES & Euclid & WFIRST & LSST   \\ \hline
BL&1.05&1.07&1.04&1.1  \\ \hline 
CGm&1.42&1.44&1.4&1.39  \\ 
CDm&1.35&1.48&1.49&1.52  \\ \hline 
TGm&1.42&1.53&1.37&1.49  \\ 
TDm&1.33&1.55&1.39&1.51  \\ \hline
\end{tabular}
\end{center}
\label{tab:infoloss}
\end{table}

\subsection{Alternative IA mitigation strategies}
\label{sec:alternative}

\begin{figure*}
\includegraphics[width=17cm]{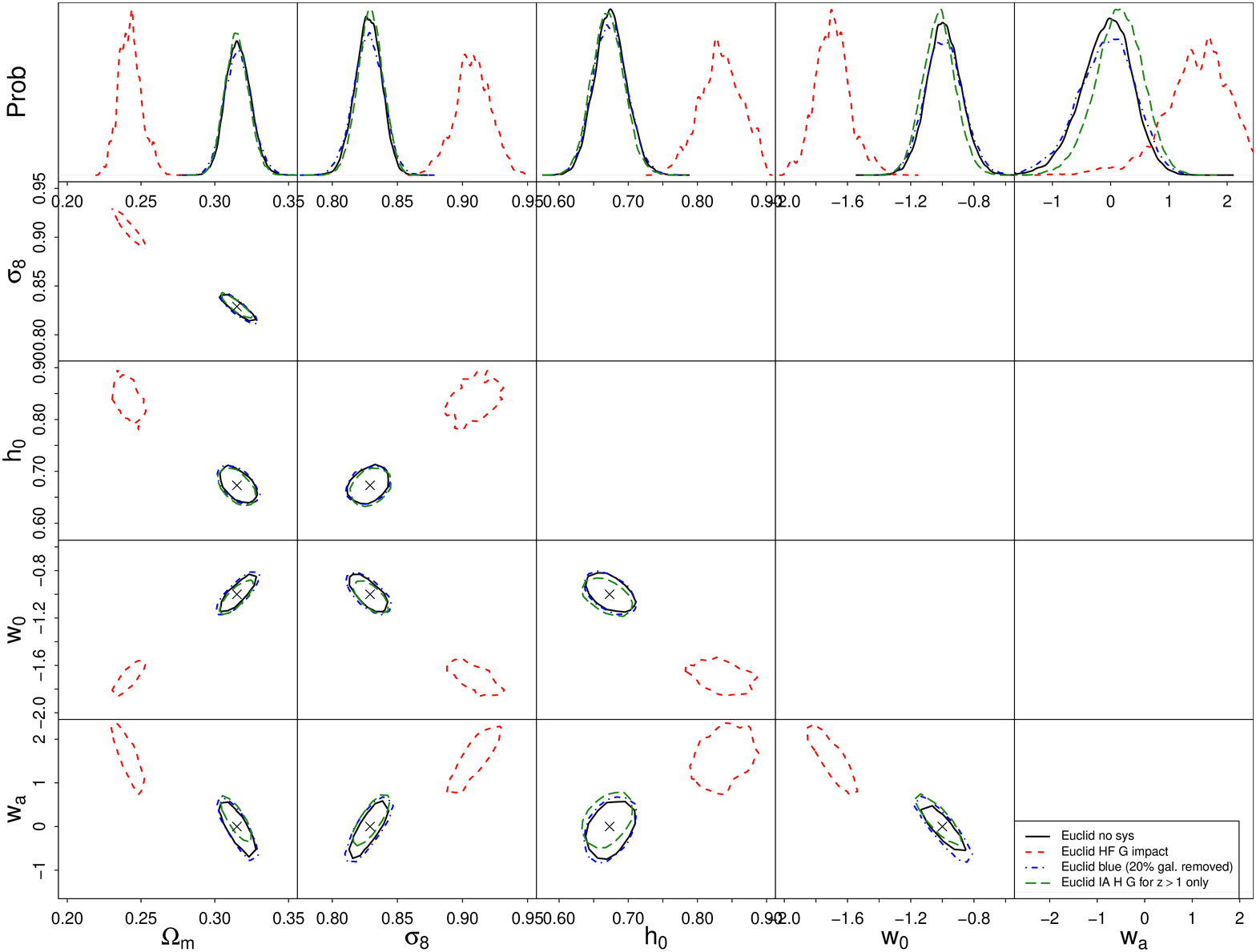}
\caption{Impact of IA (NLA Halofit, GAMA LF) for the Euclid mission (\ti{red/dashed}) compared to the Euclid statistical errors (\ti{black/solid}). The \ti{blue/dot-dashed} contours correspond to removing 20\% of the Euclid galaxies, which mimics an IA mitigation through conservatively removing the contaminated, red galaxy fraction. The \ti{green/long-dashed} contours show the impact of IA (fiducial model,$f_\mr{red}=20\%$), but with IA being completely removed for $z \le 1$, which mimics an analysis that includes an overlapping low-z spectroscopic survey. }
        \label{fi:Euclid_alter}
\end{figure*}

In this section we investigate alternative IA mitigation strategies for Euclid, since we have established that Euclid is most severely impacted by IA. Figure \ref{fi:Euclid_alter} shows the Euclid statistical errors (\ti{black/solid}) in comparison to the bias that occurs when analyzing the fiducial Euclid IA data vector without any mitigation (\ti{red/dashed}). 

Under the assumptions that (1) only red galaxies are intrinsically aligned and that (2) we can precisely distinguish between red and blue galaxies, it is an obvious mitigation strategy to remove the red fraction.  The \ti{blue/dot-dashed} contours correspond to removing 20\% of the Euclid galaxies, which results in a number density of 16/$\mr{arcmin^2}$. The constraining power is hardly diminished, which as mentioned before can be explained by the fact that the covariance is dominated by non-Gaussian terms rather than the shape noise term.   

Obviously the assumptions (1) and (2) are optimistic. Tidal torquing mechanisms cause higher-order IA effects in blue galaxies, and on sufficiently large scales, a contribution linear in the tidal field may dominate for these galaxies as well. Given the current observational upper limits on IA of blue galaxies, the effect is non-negligible and can potentially cause severe biases in cosmological parameters. Targeted spectroscopic information is needed to improve our understanding of IA of these galaxies. 

The \ti{green/long-dashed} contours correspond to an analysis where we assume an IA contaminated (red galaxies only) Euclid data vector (NLA Halofit, GAMA LF), but the IA signal is set to zero for $z \le 1$. We find that the severe bias seen in \ti{red/dashed} contours has almost completely vanished, indicating that at high redshifts the shear signal completely dominates the IA contamination. 

This result motivates a low-z spectroscopic survey (or multi-band photometric survey as proposed in \cite{SPHEREx}) that overlaps with the corresponding imaging survey to determine the red galaxy IA scale dependence and amplitude over the relevant redshift range. If a negligible level of IA from blue galaxies (consistent with current non-detections) can be confirmed with targeted spectroscopic observations, a wide, shallow ($z < 1$) spectroscopic survey may be sufficient for the calibration of IA mitigation schemes.

\section{Discussion and Conclusions}
\label{sec:disc}
In this paper we present forecasts on the impact of intrinsic alignments on future surveys and explore the performance of an IA mitigation strategy that uses the NLA model as a template, employing 10-12 nuisance parameters to account for luminosity and redshift scaling of the IA amplitude and the color and redshift distribution of source galaxies.

Compared to previous studies of the impact of IA on cosmic shear \citep[e.g.,][]{Kitching08,Joachimi10,kbs10,kirk12,kbh15}, this analysis uses more detailed and realistic models for the IA contamination, the expected statistical noise, and forecasting methods:
\begin{itemize}
\item In this study, modeling of the IA amplitude accounts for survey specific red fraction and luminosity distribution of source galaxies, while previous studies \citep[e.g.,][]{kbh15} used the amplitude normalization of the bright SuperCOSMOS galaxy sample \citep{his04, brk07} to represent the alignment of all source galaxies \citep[but see][for a recent forecast including luminosity dependent normalization]{Joachimi15}. We note that accounting for the luminosity dependence of the IA amplitude reduces the amplitude of parameter biases considerably.
\item Previous forecasts relied on Gaussian covariances while we include higher-order moments of the density field in our error bars. In particular the inclusion of Halo Sample Variance causes a significant correlation of power spectrum measurement at large $l$, which increases the statistical uncertainty and hence decreases the significance of biases.
\item Previous analyses used the Fisher forecasting formalism, while this study uses simulated likelihood analyses using MCMC. Fisher analyses inherently assume that the surface of the posterior probability is described by a Gaussian, and the systematic effects can be linearized in the nuisance parameters to calculate biases. While this is an accurate description at the best-fit parameter point, it becomes much less accurate for the outer regions of the parameter space. While it is not clear whether Fisher analyses over- or underestimate the bias (it can go either way), it is clear that an MCMC is more accurate and may show very different results, especially in high-dimensional, degenerate parameter spaces.  
\item We consider a large range of IA scenarios to encompass IA modeling uncertainties. We also consider the recently developed full tidal alignment model \cite{bvs15}, which has been found to be in excellent agreement with IA measurements. 
\end{itemize}

Based on these improved forecasting capabilities we find the unmitigated bias from IA on cosmological parameters to be less severe than most previous studies. The significance of these biases is highest for shallow surveys (as the relative IA contamination is highest at low redshift), and increases with survey area as the survey's constraining power increases. 

As described in Sect.~\ref{sec:fred}, we average the observed redshift and luminosity scaling on the IA amplitude from \citet{jma11} over the source galaxy LF. Due to the current lack of representative luminosity function measurements to the depth required for deep future surveys, we extrapolate two moderate-redshift measurements of the LF and its evolution. The difference between these two models is significant, and modeling of the galaxy LF should be considered a key uncertainty in future IA forecasts. 

We propose an IA mitigation strategy that is based on the NLA Halofit template for the IA power spectrum shape and uses 10 nuisance parameters to model the amplitude as a function of redshift and source galaxy distribution. We find this mitigation strategy to be sufficient to remove the parameter bias for the non-linear IA models considered in this study (NLA Halofit, NLA Coyote, full tidal alignment), and for all surveys (DES, Euclid, LSST, WFIRST) considered. This indicates that for the currently planned stage IV surveys IA mitigation does not require an exact model for IA in the non-linear regime; an approximate model will be sufficient.

The proposed mitigation strategy breaks down if the difference between the IA scenario contaminating the data vector and the NLA Halofit template assumed in the mitigation becomes too large, as demonstrated by the example of mitigating a linear alignment contamination with a NLA Halofit template. However, recent comparisons of observational data and IA models \citep[e.g.][]{jma11, bvs15} strongly prefer NLA and full tidal alignment models over the linear alignment model, hence this specific failure is likely less of a concern for practical purposes. However, in order to test the fixed shape assumption, one can include a simple parameterization of the IA power spectrum shape, e.g. $P_{\delta\mathcal{E}}(k,z) =P_\delta^\beta(k,z) P_{\rm{lin}}^{(1-\beta)}(k,z)$ with $\beta$ as an additional nuisance parameter, in future implementations of this mitigation scheme.

As implemented here, this mitigation strategy reduces a survey's constraining power on individual cosmology parameters by several tens of percents (c.f. Table~\ref{tab:infoloss}). We emphasize that these numbers are overly pessimistic due to the priors on IA and LF parameters employed in this analysis (see Table~\ref{tab:nuisance}). These are several times broader than current observational constraints, which was required due to the broad range of IA power spectrum models presented here. It it unclear how well these parameterization will extend to faint luminosities and high redshifts, hence measuring the IA parameters as demonstrated in \cite{mbb11,jma11}, and the LF of the source galaxy sample is the most promising way forward to further enhance IA mitigation schemes. The wide prior ranges on IA and LF parameters are also the reason why our joint analysis of IA and Gaussian photo-z uncertainties is dominated by the former. We defer an in-depth analysis of the degeneracies between IA self-calibration with stringent priors and more realistic photometric redshift uncertainties to future work. 

In general, the success of a mitigation scheme that marginalizes over the parameters of a specific model for a systematic effect will depend on whether the true signal from this effect is more degenerate with the chosen nuisance parameters (success) or the cosmological parameters (failure). If multiple systematics models are viable, one might consider a hybrid model, e.g.\ allowing the analysis code to use a weighted combination of two models, $A \times \mr{Model}_1 + B \times \mr{Model}_2$, and marginalizing over $A, B$ in addition to the nuisance parameters within $\mr{Model}_1$ and $\mr{Model}_2$. In the context of IA mitigation, this will be particularly relevant for more detailed modeling and mitigation of blue galaxy alignments. 

If the upper limits on blue IA continue to decrease with future observations, and if red and blue galaxies can be identified with sufficient accuracy, our analysis suggests removing red galaxies from the cosmic shear analysis. One interesting results of this paper is that a removal of 20\% of the galaxies only cause a loss in information of 4-10\% depending on the survey considered. Choosing a well-understood galaxy (sub)sample for future cosmic shear analyses, instead of maximizing the galaxy number density at all costs, may be a favorable strategy.

Any IA mitigation technique will be further enhanced by extending the cosmic shear analysis to a joint analysis of cosmic shear and other observables required to further self-calibrate the IA amplitude from the same data set, i.e. low-redshift measurements of the (photometric) galaxy position-shape correlation, and clustering correlation function measurements for the same galaxy sample (to self-calibrate galaxy bias parameters required for modeling the galaxy position-shape correlation). However, such a joint analysis including non-Gaussian (cross-) covariances for all observables is beyond the scope of this paper.

\section*{Acknowledgments}
The authors thank Rachel Mandelbaum for many helpful comments and discussions. This paper is based upon work supported in part by the National Science Foundation under Grant No. 1066293 and the hospitality of the Aspen Center for Physics. Part of the research was carried out at the Jet Propulsion Laboratory, California Institute of Technology, under a contract with the National Aeronautics and Space Administration. TE thanks the JPL High-Performance Computing team for outstanding support.

\bibliographystyle{mn2e}

\end{document}